\documentclass[journal,twocolumn,10pt]{IEEEtran}
\usepackage{amsmath,cite,amsfonts,amssymb,graphicx,color, mathtools,setspace,comment,float}
\usepackage[justification=centering]{caption}
\usepackage{subfigure}

\usepackage{multirow}
\usepackage{bigstrut}
\usepackage[belowskip=-13pt,aboveskip=2pt]{caption}

\begin{document}
\title{Limited Feedback Massive MISO Systems with Trellis Coded Quantization for Correlated Channels}
\author{Jawad~Mirza,~\IEEEmembership{Student Member,~IEEE,}
        Mansoor~Shafi,~\IEEEmembership{Life Fellow,~IEEE,}
        Peter~J.~Smith,~\IEEEmembership{Fellow,~IEEE,}
        and~Pawel~A.~Dmochowski,~\IEEEmembership{Senior Member,~IEEE}
        }

\maketitle

\begin{abstract}
In this paper, we propose trellis coded quantization (TCQ) based limited feedback techniques for massive multiple-input single-output (MISO) frequency division duplexing (FDD) systems in temporally and spatially correlated channels. We exploit the correlation present in the channel to effectively quantize channel direction information (CDI). For multiuser (MU) systems with matched-filter (MF) precoding, we show that the number of feedback bits required by the random vector quantization (RVQ) codebook to match even a small fraction of the perfect CDI signal-to-interference-plus-noise ratio (SINR) performance is large. With such large numbers of bits, the exhaustive search required by conventional codebook approaches make them infeasible for massive MISO systems. Motivated by this, we propose a differential TCQ scheme for temporally correlated channels that transforms the source constellation at each stage in a trellis using 2D translation and scaling techniques. We derive a scaling parameter for the source constellation as a function of the temporal correlation and the number of BS antennas. We also propose a TCQ based limited feedback scheme for spatially correlated channels where the channel is quantized directly without performing decorrelation at the receiver. Simulation results show that the proposed TCQ schemes outperform the existing noncoherent TCQ (NTCQ) schemes, by improving the spectral efficiency and beamforming gain of the system. The proposed differential TCQ also reduces the feedback overhead of the system compared to the differential NTCQ method. 

\end{abstract}


\IEEEpeerreviewmaketitle
\vspace{-5pt}
\section{Introduction}

\IEEEPARstart{M}{assive} multiple-input single-output (MISO) cellular systems use large numbers of transmit antennas at a base station (BS) to simultaneously serve a smaller number of users \cite{rusek2013scaling}. This results in a higher spectral efficiency, less inter-user interference and reduced energy consumption \cite{6736761,rusek2013scaling,marzetta2010noncooperative}. The use of large numbers of antennas at the BS also provides highly directional beamforming \cite{brunobook} and the array gain from beamforming can improve the link budget. Due to these and several other attractive features, massive MISO is becoming a popular contender for 5G wireless systems. However, there are a number of factors that limit the performance of massive MISO systems, e.g. pilot contamination \cite{5898372,marzetta2010noncooperative}, reduced multiuser (MU) diversity gain due to channel hardening \cite{1327795} and high spatial correlation at the BS \cite{call}. An overview of massive
multiple-input multiple-output (MIMO) is described in \cite{rusek2013scaling} including information theoretic aspects and linear transceivers along with the main design features and practical challenges. The attractive features of massive MIMO systems also apply to massive MISO systems.

The expressions for signal-to-noise ratio (SNR) and signal-to-interference-plus-noise ratio (SINR) for massive MU MISO systems are derived in \cite{rusek2013scaling} for both zero-forcing (ZF) and matched-filter (MF) precoding schemes, respectively. At high SNR, ZF precoding gives superior performance compared to MF precoding because the interference is strong and ZF cancels the interference whereas MF only maximizes the desired signal. In \cite{yang2013performance}, it is observed that if a higher spectral efficiency is required, ZF precoding is preferable in the high SNR region, while MF (also known as conjugate beamforming) precoding is preferable in terms of energy efficiency. Analytical spectral efficiency approximations are derived in \cite{hoydis2013massive} with several linear precoders and detectors for non-cooperative multi-cell massive MISO systems using time-division duplexing (TDD) operation. There are also several other studies that deal with the performance of massive MIMO/MISO systems with linear precoders \cite{6377481,wagner2012large,6241389,gao2011linear}. In MU massive MISO systems, the computational complexity increases with large numbers of antennas and users. This increase causes delays in learning the channel estimate at the BS, resulting in outdated precoders, especially when the channel is changing rapidly over time. To overcome this ``channel aging'', a channel prediction method has been proposed in \cite{6608213}.

In TDD transmission, the BS acquires downlink channel state information (CSI) via uplink training and exploiting channel reciprocity, whereas, in frequency division duplexing (FDD) operation, this is achieved via a low-rate feedback link. Although, most of the research on massive MISO consider TDD transmission, most of the existing cellular systems use FDD operation. In order to equip the transmitter with CSI, TDD based MISO systems require tight RF synchronization between transmit and receive antennas \cite{1404883}. However, if there are downlink/uplink  synchronization errors in RF chains then the channel estimated at the transmitter using reciprocity may not be precise \cite{choi2014,1404883}. In this paper, like many other studies \cite{choi2013,choi2014,kuo2012compressive,au2011trellis}, we investigate massive MISO systems with FDD operation.

The feedback overhead is large in massive MISO systems. Conventional codebook-based limited feedback schemes, discussed for independent and identically distributed (i.i.d.) Rayleigh fading channels in \cite{love2003grassmannian,love2008overview,au2007performance} and for correlated channels in \cite{raghavan2007systematic,kim2011mimo,choi2012new,love2006limited,dMirzaLimited,6648514,6824239}, are not feasible as the number of codewords required in the codebook grows exponentially with the number of transmit antennas, making a search for an appropriate codeword a computationally complex task. There are very few studies that explore limited feedback schemes for massive MISO FDD systems that reduce the computational complexity of the search for an appropriate codeword at a user. 

For a spatially correlated channel, a compressive sensing-based feedback
scheme is proposed in \cite{kuo2012compressive}, where the feedback contents are dynamically configured depending on channel conditions. Recently, open-loop and closed-loop training techniques have been proposed in \cite{choi2014} for massive MISO FDD systems, where long-term channel statistics and previously received training signals are used to increase the performance of channel estimation at each user. Moreover, with a small amount of feedback overhead, it is shown in \cite{choi2014} that the closed-loop training scheme reduces the downlink training overhead. Recently, trellis based channel quantization methods have drawn much attention in limited feedback massive MISO FDD systems due to their low complexity compared to conventional codebook approaches.

In trellis coded quantization (TCQ) based limited feedback systems, the channel is quantized by the user using the concept of a trellis with a source constellation and the Viterbi algorithm. The input bit sequence of the selected trellis path is then fed back to the BS, where the BS uses a convolutional coder to regenerate the quantized channel and compute the beamforming vectors. In conventional channel coding, the Viterbi algorithm is run over the trellis stages where each stage represents time. In contrast, in limited feedback MISO systems with TCQ, each trellis stage corresponds to a particular BS antenna element. One way of selecting the trellis path is by using brute-force maximum-likelihood (ML) optimization that searches for the best path (most likely sequence) from the set of all the trellis paths (also known as trellis codewords). However, due to the size of exhaustive search, it is not practical to consider brute-force ML, therefore the Viterbi algorithm is preferred as it reduces the search complexity by relying on the surviving trellis paths only. 

A trellis based channel quantization scheme is proposed in \cite{au2011trellis} for single-user (SU) multi-cell MISO systems with large numbers of BS antennas. The main idea is to use the TCQ scheme \cite{marcellin1990trellis} to quantize the channel at each user using Ungerboeck's trellis coded modulation (TCM) approach \cite{ungerboeck1982channel} and the Viterbi algorithm. A noncoherent TCQ (NTCQ) approach for a massive MISO system is proposed in \cite{choi2013}, where a bank of coherent detectors is implemented to realize near optimal noncoherent detection. Here, TCQ with Ungerboeck's trellis is used to quantize the channel direction information (CDI) for a SU massive MISO system. By adopting  Ungerboeck's TCM structure, the TCQ scheme uses source constellations such as QPSK, 8PSK or 16QAM to quantize each channel entry with 1 bit, 2 bits or 3 bits, respectively (this will be clarified for the QPSK constellation in Section \ref{4a}). The Viterbi algorithm \cite{forney1973viterbi} is used to search for the optimal path in a trellis and a convolutional code is used at the BS to reconstruct the quantized channel, using the trellis sequence as input and producing the quantized channel vector at the output. Three different limited feedback schemes are proposed in \cite{choi2013} for three different channel models; an i.i.d. Rayleigh fading channel, a temporally correlated channel and a spatially correlated channel. In \cite{choi2013}, the quantization process of the temporally correlated channel requires additional feedback of optimization parameters, hence increasing the feedback overhead. The adaptive TCQ method proposed in \cite{choi2013} for spatially correlated channels requires knowledge of the transmit correlation at the transmitter.

In this paper, we adopt the TCQ framework developed in \cite{marcellin1990trellis} to quantize the MISO channel at each user for temporally correlated channels. Exploiting the temporal correlation in the channel, we design a differential scheme that transforms the source constellation at each stage in a trellis, such that it is centered around the previously selected constellation point, for the next feedback instance. We rely on 2D translation and scaling schemes to transform the source constellation. 

We also propose a TCQ based limited feedback scheme for spatially correlated channels\footnote{Similar to \cite{choi2013}, we also propose two separate TCQ schemes for temporally and spatially correlated channels in Sections \ref{ourr} and \ref{our_sp}, respectively.} where we quantize the correlated channel directly at the receiver. Hence, unlike \cite{choi2013}, the proposed scheme does not require knowledge of the transmit correlation matrix at the transmitter. For spatially correlated channels, we consider uniform linear array (ULA) and uniform rectangular array (URA) antenna topologies.

The main contributions of this paper are summarized below:
\begin{itemize}
\item To motivate the TCQ based limited feedback approach, we quantify the codebook dimension required for an ordinary limited feedback approach. In particular, we derive an expected SINR approximation for MF precoding with random vector quantization (RVQ) codebook CDI. This expected SINR approximation is used to derive the number of feedback bits required to achieve a mean SINR performance with RVQ codebooks which is $z$ dB below the mean SINR with the perfect CDI.

 \item We propose a differential TCQ method that uses a transformed source constellation at each stage of the trellis to quantize the MISO CDI. By efficiently utilizing the temporal correlation information and successively transforming the source constellation, the proposed method reduces the feedback overhead and boosts the performance of the MU massive MISO system compared to the existing differential NTCQ scheme \cite{choi2013}. 

 \item In order to track the temporally correlated channel, the proposed method uses 2D translation and scaling techniques to transform the source constellation after each feedback interval. We derive an expression for the source constellation scaling parameter, as a function of the temporal correlation and the number of BS antennas.
 
 \item We also propose a TCQ based limited feedback technique for spatially correlated channels where the correlated channel is quantized at the receiver directly without the need to decorrelate it. The advantage of the proposed scheme is that the transmitter reconstructs the channel without requiring any knowledge of the spatial correlation.
\end{itemize}

Additionally, we also compare two commonly used linear precoding schemes, namely ZF and MF and evaluate their performance in terms of mean values of the  SINR and spectral efficiency.
The NTCQ scheme \cite{choi2013} serves as a baseline for this study. Although the NTCQ method was originally proposed for a SU massive MISO system, it can also be used in a MU setting. In \cite{choi2013}, multiple Viterbi algorithms run in parallel, each searching for the best output path over different values of amplitude scalings and phase rotations. According to \cite{choi2013}, a parallel search in Euclidean space to quantize a channel vector is approximately equal to a quantization on the Grassmannian manifold. Moreover, due to the presence of parallel Viterbi blocks, the overall process is described as noncoherent TCQ. However, the mean beamforming gain due to the parallel set of Viterbi algorithms does not improve significantly compared to having a single Viterbi block, especially when using higher source constellations (see Fig. 8 of \cite{choi2013}). Like the NTCQ scheme, the proposed differential TCQ scheme can also be implemented using parallel blocks of Viterbi algorithm, however, for simplicity and reduced complexity we only rely on a single Viterbi block at each user. 
 
The source constellation plays an important role in deciding the required number of feedback bits. If the source constellation consists of $L$ symbols or constellation points, then the length of the input bit sequence (or the number of feedback bits) after TCQ becomes $M\left(\log_2(L)-1\right)$, where $M$ is the number of BS antennas. Therefore, in order to keep the feedback overhead reasonable, we use QPSK  and 8PSK source constellations, requiring $M$ and $2M$ feedback bits, respectively. Another benefit is that all the constellation points in a PSK constellation are positioned with uniform angular spacing around a circle, hence making it a suitable candidate for the proposed schemes in Section \ref{ourr} and \ref{our_sp}.

\emph{Notation:} We use $(\cdot)^H$, $(\cdot)^T$ and $(\cdot)^{-1}$ to denote the conjugate transpose, the transpose and the inverse operations respectively. $\|\cdot\|$ and $|\cdot|$ stand for vector and scalar norms respectively. $\mathbb{E}[\cdot]$ denotes expectation. Bold uppercase and lowercase letters are used to represent matrices and vectors. Lowercase italic letters denote elements
of vectors/matrices.
\vspace{-5pt}
\section{Downlink System Model}\label{sm}
Consider a single-cell MU MISO system with $M$ transmit antennas at the BS. The BS serves $K$ single antenna users simultaneously using a suitable precoding technique, where $K < M$ with a constant ratio $q= M/K$. On the downlink, the received signal for the $k^{\textrm{th}}$ user can be written as
\begin{equation}
 y_k = \mathbf{h}_k \mathbf{x} + n_k, \ \ \ \ \ \ k=1, \ldots, K,
\end{equation}
where $\mathbf{h}_k \in \mathbb{C}^{1 \times M}$ denotes the channel of the $k^{\textrm{th}}$ user and $n_k$ is the noise assumed to be i.i.d. with $n_k \sim \mathcal{CN}(0,1)$ $\forall k$. We assume uniform power allocation among $K$ users. Denoting SNR by $\rho$, the transmitted signal is given by $\mathbf{x} = \sqrt{\frac{\rho}{K}}\sum_{k=1}^K \mathbf{w}_k s_k$, where $s_k$ and $\mathbf{w}_k$ are the data symbol and $M \times 1$ unit-norm precoding vector for the $k^{\textrm{th}}$ user, respectively. The data symbols are assumed i.i.d. with $\mathbb{E}\left[|s_k|^2\right] = 1$. The received signal can be expressed as
\begin{equation} \label{recs}
 y_k = \underbrace{\sqrt{\frac{\rho}{K}} \left(\mathbf{h}_k \mathbf{w}_k \right) s_k}_\text{signal} + \underbrace{\sum_{j \neq k} \sqrt{\frac{\rho}{K}} \left( \mathbf{h}_k \mathbf{w}_j \right) s_j}_\text{interference} + \underbrace{n_k}_\text{noise}.
\end{equation}
From \eqref{recs}, the SINR of the $k^{\textrm{th}}$ user is defined as 

 \begin{equation}\label{111}
 \textrm{SINR}_k = \frac{ \frac{\rho}{K} \left| \mathbf{h}_k \mathbf{w}_k \right|^2 }
 {\frac{\rho}{K} \sum_{j \neq k}^{K} \left| \mathbf{h}_k \mathbf{w}_j \right|^2 +1}.
\end{equation}

Treating interference as noise, the expected spectral efficiency for the MU MISO system is given by \cite{jindal2006mimo}
\begin{align}
  \mathbb{E} \left[R_{\textrm{sum}}\right]  =  \mathbb{E}\left[ \sum_{k=1}^K \log_2 \left(1 + \textrm{SINR}_k \right)\right]. \label{12}
\end{align}

In this paper, we use a differential TCQ scheme (discussed in Section \ref{ourr}) to quantize the CDI vector, $\bar{\mathbf{h}}_k = \frac{\mathbf{h}_k}{\| \mathbf{h}_k\|}$, for each user. We denote the ``quantized'' CDI for the $k^{\textrm{th}}$ user by $\tilde{\mathbf{h}}_k$.
We next discuss the ZF \cite{yoo2006optimality} and MF \cite{1468466,godara1997application} linear precoding schemes that are generally considered in MU massive MISO systems. 
\vspace{-10pt}
\subsection{ZF precoding}\label{s2a}
In ZF precoding, the perfect CDI vectors, $\bar{\mathbf{h}}_k$, of all users are concatenated into a single $K \times M$ matrix at the BS, denoted by $\mathbf{H}=[\bar{\mathbf{h}}_1^T \ldots \bar{\mathbf{h}}_K^T]^T$ \cite{yoo2006optimality}. The precoding vector, $\mathbf{w}_k=\mathbf{v}_k$, is the normalized $k^{\textrm{th}}$ column of the matrix $\mathbf{V}$, where $\mathbf{V} = \mathbf{H}^H(\mathbf{H} \mathbf{H}^H)^{-1}$, such that $\mathbf{v}_k = \mathbf{V}(:,k)/ \|\mathbf{V}(:,k)\|$. With perfect CDI, ZF precoding completely eliminates interference and using the results given in \cite{peel2005vector,Hochwald02space-timemultiple,rusek2013scaling} for large $M$, the expected SINR can be approximated as
\begin{equation}\label{tt}
 \mathbb{E}\left[\textrm{SINR}_k^{\textrm{ZF}}\right] \approx \frac{\rho}{K} \mathbb{E}\left[\left|\mathbf{h}_k \mathbf{w}_k \right|^2\right] = \rho \left(q-1\right).
\end{equation}

Unlike perfect CDI, with limited feedback based quantized CDI, $\tilde{\mathbf{h}}_k$, ZF precoding does not eliminate the interference completely due to quantization errors.
\vspace{-10pt}
\subsection{MF precoding}
For MF precoding with perfect CDI, we have $\mathbf{w}_k = \bar{\mathbf{h}}_k^H$ and the expected SINR approximation for the $k^{\textrm{th}}$ user, derived in the Appendix, is given by
\begin{align}
\mathbb{E} \left[\textrm{SINR}_k^{\textrm{MF}}\right] \approx \frac{ \rho q }
{\frac{\rho \left(K-1 \right)}{K}+1}.\label{3}
\end{align}
At high SNR ($\rho \to \infty$), we can write \eqref{3} as
\begin{equation}
\lim_{\rho \to \infty} \mathbb{E} \left[\textrm{SINR}_k^{\textrm{MF}}\right] \approx \frac{M}{K-1}.\label{33}
\end{equation}
In contrast to \eqref{tt}, it is evident from \eqref{33} that there is no improvement in the MF SINR in the limit as $\rho$ increases, hence limiting the SINR gain. 

\vspace{-10pt}
\section{Impracticality of RVQ codebooks}\label{RVQ}
In this section, we show that codebook based limited feedback approaches are not practical in massive MISO systems. Although this is known in an intuitive sense, here we quantify the scale of the problem by computing the size of the codebook required. A similar analysis is derived in \cite{jindal2006mimo} for ZF precoding where the spectral efficiency loss is investigated. In contrast, we consider MF precoding with RVQ codebooks and investigate the loss in SINR performance. In particular, we derive the number of bits required to achieve an expected SINR performance with RVQ codebooks that suffers a $z$ dB loss compared to the expected SINR performance with perfect CDI. We also briefly discuss the search complexity of the quantization process with RVQ codebooks. Such calculations require the RVQ assumption but give a clear indication of the order of the codebook size for other more practical schemes.

Consider a limited feedback system where the CDI is quantized using an RVQ codebook of size $N_c$, thus requiring $b=\log_2 (N_c)$ feedback bits per user. Let us denote the selected RVQ codeword vector of size  $M \times 1$ for the $k^{\textrm{th}}$ user as $\tilde{\mathbf{h}}_k$. For MF precoding with RVQ limited feedback, the expected SINR approximation for the $k^{\textrm{th}}$ user, denoted by $\mathbb{E}\left[\overline{\textrm{SINR}}_k^{\textrm{MF}}\right]$, can be approximated using the approach given in \cite{5510182} as
\begin{align}
\mathbb{E}\left[\overline{\textrm{SINR}}_k^{\textrm{MF}}\right] &\approx \frac{ \frac{\rho}{K}  \mathbb{E} \left[ | \mathbf{h}_k \tilde{\mathbf{h}}_k^H |^2 \right]} {\frac{\rho}{K} \mathbb{E} \left[ \sum_{j \neq k}^{K} | \mathbf{h}_k \tilde{\mathbf{h}}_j^H |^2 \right]+1} \label{5} \\
&=\frac{ \frac{\rho}{K} \mathbb{E}\left[\|\mathbf{h}_k\|^2\right] \mathbb{E} \left[ | \bar{\mathbf{h}}_k \tilde{\mathbf{h}}_k^H |^2 \right]} {\frac{\rho}{K} \mathbb{E}\left[\|\mathbf{h}_k\|^2 \right]  \sum_{j \neq k}^{K} \mathbb{E} \left[| \bar{\mathbf{h}}_k \tilde{\mathbf{h}}_j^H |^2 \right]+1},\label{5r}
\end{align}
where the bar denotes the SINR resulting from the limited feedback RVQ CDI. Equation \eqref{5r} comes from the independence between the amplitude and direction of $\mathbf{h}_k$. It is shown in \cite{au2007performance}, that the expectation $\mathbb{E} \left[ | \bar{\mathbf{h}}_k \tilde{\mathbf{h}}_k^H |^2 \right]$ is given by 
\begin{equation}\label{6}
 \mathbb{E} \left[ | \bar{\mathbf{h}}_k \tilde{\mathbf{h}}_k^H |^2 \right] = 1 - N_c B\left(N_c,\frac{M}{M-1} \right) \overset{\Delta}{=} 1 - \xi,
\end{equation}
where $B(\cdot,\cdot)$ denotes a beta function. The quantity $\xi$ represents the expected value of quantization errors. In \cite{jindal2006mimo}, an upper bound on $\xi$ is given by
\vspace{-5pt}
\begin{align}
 \xi &\leq  2^{\frac{-b}{M-1}}. \label{7}
 \end{align}
Due to the independence between the unit norm vectors $\bar{\mathbf{h}}_k$ and $\tilde{\mathbf{h}}_j$, the second expectation in the denominator of \eqref{5r}, is $\mathbb{E} \left[ | \bar{\mathbf{h}}_k \tilde{\mathbf{h}}_j^H |^2 \right] = \frac{1}{M}$ and $\mathbb{E}\left[\|\mathbf{h}_k\|^2 \right] = M$. Substituting these expectations and \eqref{6} into \eqref{5r} gives the result
\begin{equation}\label{9}
\mathbb{E}\left[\overline{\textrm{SINR}}_k^{\textrm{MF}}\right] \approx \frac{\rho q \left(1 - \xi \right)}{ \frac{\rho}{K} \left( K-1\right)+1}.
\end{equation}
At high SNR $(\rho \to \infty)$, \eqref{9} becomes
\begin{equation}\label{10}
 \lim_{\rho \to \infty} \mathbb{E}\left[\overline{\textrm{SINR}}_k^{\textrm{MF}}\right] \approx \frac{M \left(1 - \xi \right)}{ K-1},
\end{equation}
since $M=qK$. The approximation for MF precoding with RVQ codebooks allows us to compute the number of bits required to achieve an expected SINR with RVQ which is $z$ dB below the expected SINR with perfect CDI, i.e., 
\begin{align}
\frac{\mathbb{E}\left[\textrm{SINR}_k^{\textrm{MF}}\right]}{10^{\frac{z}{10}}} = \mathbb{E}\left[\overline{\textrm{SINR}}_k^{\textrm{MF}}\right].\label{99r}
\end{align}
Substituting \eqref{3} and \eqref{9} in \eqref{99r} gives
\begin{align}
\frac{ \rho q }{10^{\frac{z}{10}} \left( \frac{\rho \left(K-1 \right)+1 }{K}\right)} = \frac{\rho q \left(1 - 2^{\frac{-b}{M-1}} \right)}{\frac{\rho}{K} \left( K-1\right)+1}.\label{36}
\end{align}
Using \eqref{36} and solving for the number of feedback bits required, we have
\begin{equation}\label{rv}
 b_{\textrm{req}}^z = - (M-1) \log_2 \left( 1- 10^{-\frac{z}{10}} \right).
\end{equation}
Equation \eqref{rv} allows us to determine the number of bits required by RVQ codebooks to match the perfect CDI expected SINR performance with a $z$ dB loss. We note that in MF precoding systems, unlike ZF precoding \cite{jindal2006mimo}, the number of feedback bits required in \eqref{rv} does not depend on $\rho$. If $M=100$ and $z=3$ dB, the number of bits required to obtain a half of the perfect CDI expected SINR performance is, $b_{\textrm{req}}^z = 99$ bits  i.e. a codebook of size $2^{99}= 6.3383 \times 10^{29}$. Even to achieve a very low target where the signal power is equal to the interference-plus-noise power, i.e. $\mathbb{E}\left[\overline{\textrm{SINR}}_k^{\textrm{MF}}\right] \to 1$, the number of bits required is
\begin{equation}\label{37}
 b_{\textrm{req}} = - (M-1) \log_2 \left( 1- \frac{K-1 }{M} \right).
\end{equation}
This is found by solving \eqref{36} with $z=0$ and 1 on the right hand side. 
For example, for $M=100$ and $K=10$, using \eqref{37} we have $b_{\textrm{req}} = 13.4701 \approx 14$, which corresponds to 16384 codebook entries per user. While the feedback overhead with large $M$ may be acceptable, the search for an appropriate codeword within the codebook is very challenging and becomes computationally infeasible as $M$ increases. The search complexity for the RVQ codebook quantization, given by $O(M2^{BM})$, grows exponentially with large $M$. Therefore, codebook-based limited feedback schemes are infeasible for massive multi-antenna systems. This serves as a motivation to seek a non-codebook approach for limited feedback MISO systems.
\vspace{-5pt}
\section{Limited feedback with TCQ}\label{4a}
In this section, we review basic concepts of TCQ based limited feedback MISO systems. Limited feedback schemes based on TCQ are recently gaining attention in massive MISO FDD systems due to their reduced complexity compared to the conventional codebook approaches in searching for an appropriate codeword. 

The quantization of CDI at the user is performed using TCQ, which consists of two key components; a trellis and a source constellation. At the user, a trellis path that gives the minimum squared Euclidean distance to the CDI is selected using the Viterbi algorithm. 
The input bit sequence corresponding to the selected path is then fed back to the BS using a low-rate feedback link. At the BS, a convolutional coder is implemented to decode the input bit sequence of the path to obtain the corresponding output sequence. The output sequence is then mapped to the source constellation to obtain the quantized CDI at the BS. Thus, the decoder and encoder of the TCM scheme are used, respectively, to quantize and reconstruct a channel vector in the TCQ method. This is in contrast to the role of decoder/encoder in a traditional TCM. In this study, we rely on Ungerboeck's trellis structure and the corresponding convolutional coder. 

\begin{figure}[!t]
\centering
\scalebox{.37}
{\includegraphics[width=24cm, height=11cm]{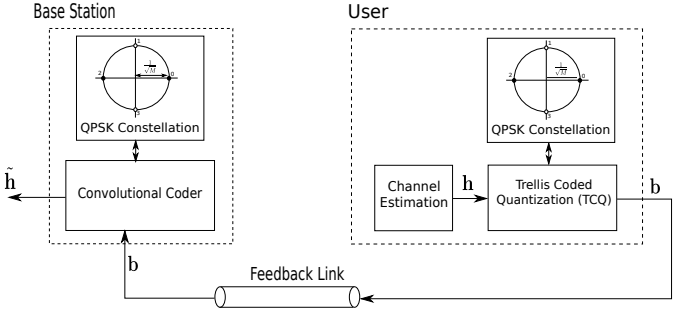}}\\
  \caption{The block diagram of the TCQ feedback process with QPSK constellation.}
  \label{Fig00}
  \end{figure}
The block diagram of the feedback process is shown in Fig. \ref{Fig00}, where the perfectly estimated channel, $\mathbf{h}$, is quantized at a user with the $N$-state trellis decoder and source constellation, using the Viterbi algorithm. The complexity of the Viterbi algorithm is $O(LNM)$, where $L$ is the total number of points in the source constellation and $N$ is the number of states in the trellis. Note that, before implementing TCQ, the channel vector is normalized to obtain CDI, such that, $\bar{\mathbf{h}}=\mathbf{h}/\|\mathbf{h}\|$. After implementing the TCQ, the input bit sequence, $\mathbf{b}$, (of length $M$) of the selected path is fed back to the BS, where it is decoded to recover the corresponding output sequence of length $2M$, using a convolutional coder. The output bit sequence is then mapped onto the source constellation to reconstruct the quantized channel vector, $\mathbf{\tilde{h}}$. 
 \begin{figure}[!b]
    \centering
        \scalebox{.32}
     {\includegraphics[width=8cm, height=8cm]{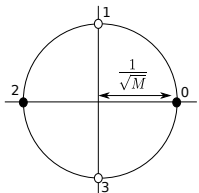}}
      \caption{The normalized QPSK constellation points.}\label{fig:left}
    \end{figure}
      \begin{figure}[!t]
    \centering
      \scalebox{.35}
     {\includegraphics[width=19cm, height=10cm]{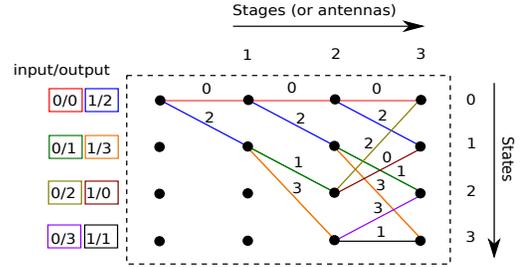}}
      \caption{The 4-state, rate 1/2 Ungerboeck trellis structure.}\label{fig:right}
  \end{figure} 
Consider an example with the QPSK constellation along with a trellis structure having $N = 4$ states. The QPSK constellation points are normalized by the number of transmit antennas, $M$. The normalized QPSK constellation and the 4-state, rate 1/2 Ungerboeck trellis structure are shown in Fig. \ref{fig:left} and Fig. \ref{fig:right}, respectively. The decimal numbers 0, 1, 2 and 3 (or 00, 01, 10, and 11 in binary) represent QPSK constellation points in Fig. \ref{fig:left}. In Fig. \ref{fig:right}, there are only two state-transitions from any given state. Each transition is mapped to a single QPSK point, hence each channel entry of $\bar{\mathbf{h}}$ will be quantized with one of the two BPSK sub-constellations represented by black and white circles in Fig. \ref{fig:left}.

The main idea in TCQ encoding is to advance through an N-state trellis, where the $m^{\textrm{th}}$ stage corresponds to the $m^{\textrm{th}}$ antenna channel. At any particular stage, there will be only $N$ survivor paths in the Viterbi algorithm. We label the paths by their respective output symbols. For example, starting from the state 0 and moving through all the paths in the trellis to reach stage 3, gives $2N$ total paths. At stage 3, each state will have two paths terminating at it. As illustrated in Fig. \ref{fig:right}, we have the following paths: [0,0,0] and  [2,1,2] at state 0, [0,0,2] and [2,1,0] {at state 1}, [0,2,1] and [2,3,3] at state 2, and [0,2,3] and [2,3,1] at state 3.

\begin{figure}[!t]
\vspace{1em}
\centering
\scalebox{.4}
{\includegraphics[width=21cm, height=4.8cm]{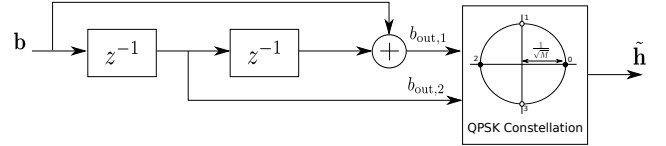}}\\
  \caption{The convolutional coder corresponding to the 4-state rate 1/2 Ungerboeck trellis. In this figure, $\mathbf{b}$ is the input bit sequence, while $b_{out,1}$ and $b_{out,2}$ are the corresponding output bits for each entry of the input.}
  \label{Fig01}
\end{figure}

The path $\mathbf{p}_{2}=[2,1,2]$ terminating at state 0 corresponds to the output vector, $\textrm{out}(\mathbf{p}_{2}) = \left[\frac{-1}{\sqrt{M}},\frac{+j}{\sqrt{M}},\frac{-1}{\sqrt{M}}\right]$ from the QPSK constellation in Fig. \ref{fig:left} and the input bit sequence is $\mathbf{b}=[1,0,0]$. The user selects the best path from each state that gives minimum squared Euclidean distance to the normalized channel vector $\bar{\mathbf{h}}$. The path metric can be defined at the $m^{\textrm{th}}$ stage as \cite{choi2013}
\begin{align}
 \textrm{metric}(\mathbf{p}^{(m)}) = \| \bar{\mathbf{h}}^{(m)} - \textrm{out}(\mathbf{p}^{(m)})  \|_2^2. \label{p0}
 \end{align}
where $\bar{\mathbf{h}}^{(m)}$ is a truncated normalized channel vector up to the $m^{\textrm{th}}$ channel entry. Equation \eqref{p0} can also be written recursively as \cite{choi2013}
 \begin{align}
 \textrm{metric}(\mathbf{p}^{(m)}) =  \textrm{metric}(\mathbf{p}^{(m-1)}) + | \bar{h}^{(m)} - \textrm{out}(p^{(m)})  |^2, \label{p1}
\end{align}
where $\bar{h}^{(m)}$ and $p^{(m)}$ are the $m^{\textrm{th}}$ entries of $\bar{\mathbf{h}}$ and $\mathbf{p}^{(m)}$. The solution to \eqref{p1} is obtained via a Viterbi algorithm that minimizes the path metric. This enables us to determine the quantized CDI for large antenna numbers in a piecewise manner.
Figure \ref{Fig01} shows the convolutional coder corresponding to the 4-state rate 1/2 Ungerboeck trellis, used to reconstruct the quantized channel at the BS.

In this paper, we assume that the TCQ scheme always starts from state 0. Therefore, it does not require the additional $\log_2(N)$ bits to be fed back to the BS indicating the starting state. Due to large channel dimensions, we are dealing with a long trellis structure, therefore the quantization errors associated with always starting from state 0 are not significant.  

\begin{figure}[!t]
  \centering
  \hspace{4pt}
  \subfigure[8-state rate 2/3 Ungerboeck trellis structure for 8PSK constellation.]{%
    {\includegraphics[width=5.5cm,height=4cm]{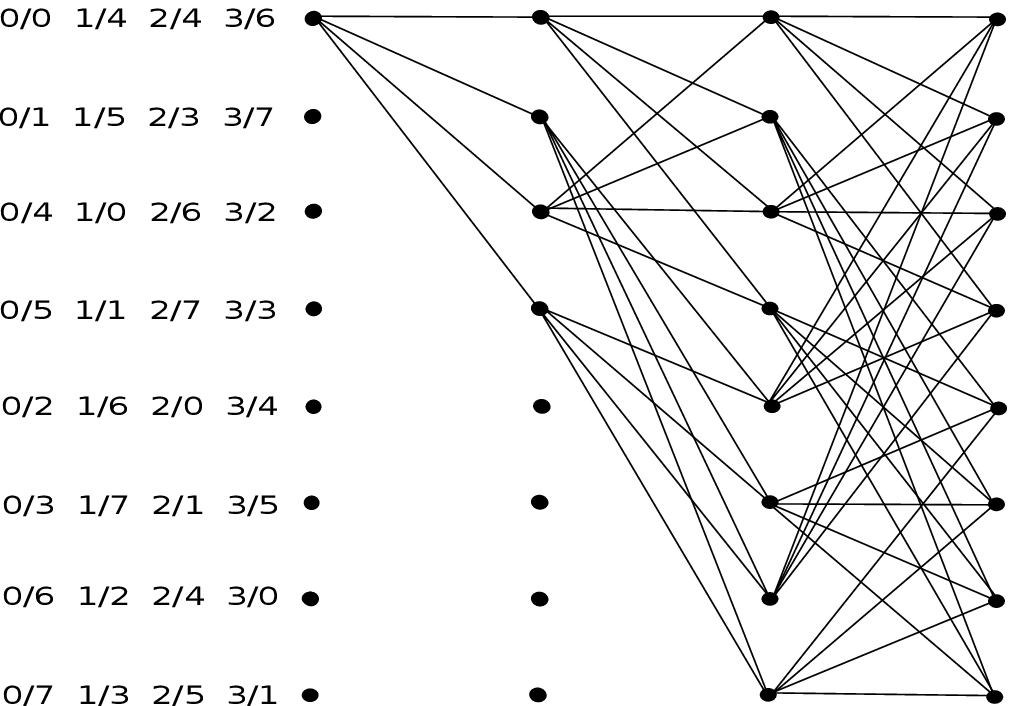}}
    \label{fig:c1}%
  }\\
  \hspace{-4pt}
   \subfigure[Convolutional coder corresponding to 8-state rate 2/3 Ungerboeck trellis structure for 8PSK constellation.]{%
    \includegraphics[width=0.37\textwidth]{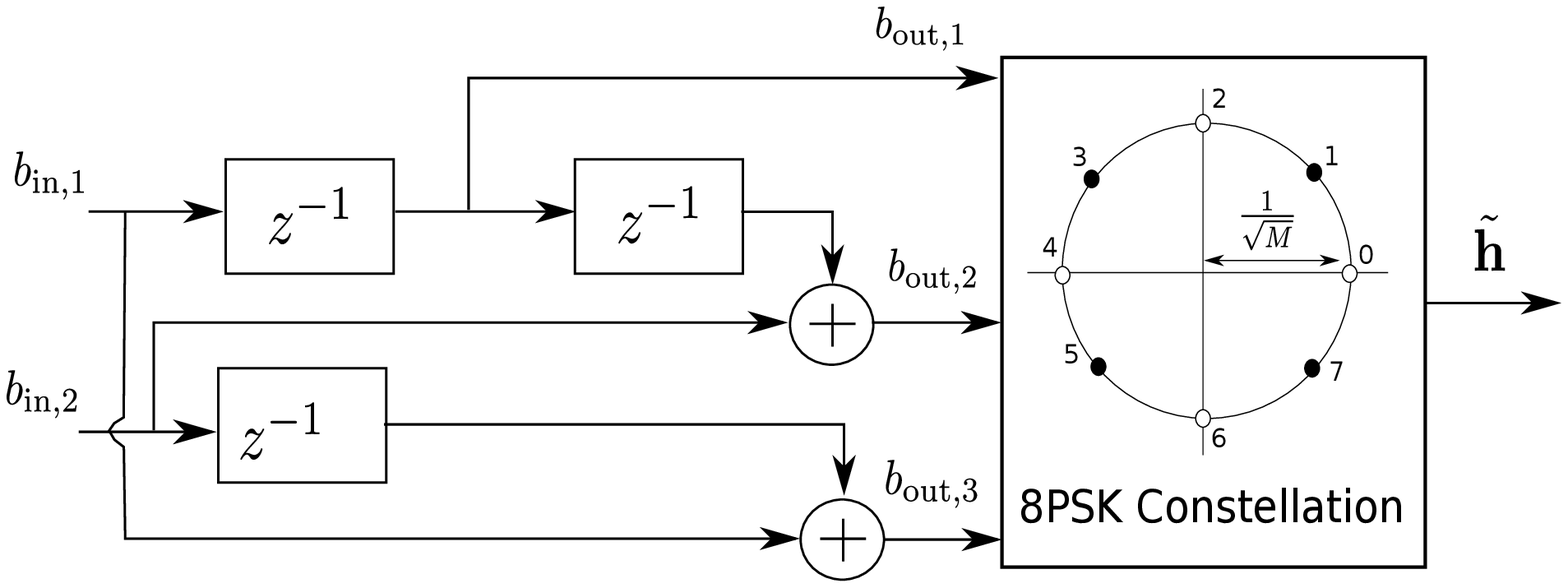}%
    \label{fig:c2}%
    }
  \caption{Trellis structure and convolutional coder for limited feedback TCQ with 8PSK constellation.}
  \label{cc}
\end{figure}

The performance of the TCQ based limited feedback scheme depends on the constellation size with larger constellations giving smaller quantization errors. However, using higher order source constellations increases the feedback overhead. In this paper, we also consider an 8PSK source constellation that consists of $L=8$ constellation points. The 8-state, rate 2/3 Ungerboeck trellis structure and corresponding convolutional coder are shown in Fig. \ref{fig:c1} and Fig. \ref{fig:c2}, respectively. 
The limited feedback TCQ method with an 8PSK constellation is similar to QPSK constellation, where, with the 8PSK constellation, each channel entry is quantized by 2 bits.
\section{Proposed Differential TCQ for Temporally correlated channels}\label{ourr}
In this section, we present the proposed differential TCQ method for temporally correlated channels. The temporally correlated channel is modeled by a first-order Gauss-Markov process where the channel of the $k^{\textrm{th}}$ user at time $t$ is given by a $1 \times M$ vector

\begin{equation}\label{0}
 {\mathbf{h}}_{k}[t] = \epsilon {\mathbf{h}}_{k}[t-1] + \sqrt{1-\epsilon^2} \mathbf{g}_{k}[t],
\end{equation}
where ${\mathbf{h}}_{k}[t]$ and ${\mathbf{h}}_{k}[t-1]$ are the current and previous channel vectors for the $k^{\textrm{th}}$ user, such that $\mathbb{E}[\|{\mathbf{h}}_{k}[t]\|^2] = M$, $\mathbf{g}_{k}[t]$ is a $1 \times M$ i.i.d. $\mathcal{CN}(0,1)$ vector and ${\mathbf{h}}_{k}[0]$ is independent of the $\mathbf{g}_{k}[t]$. The time correlation coefficient is denoted by $\epsilon$, where $0 \leq \epsilon \leq 1$. The channel is highly correlated when $\epsilon \to 1$, whereas when $\epsilon \to 0$, the temporal correlation vanishes and the channel entries become independent over time. 

The proposed technique uses the TCQ method discussed in the previous section, to quantize the temporally correlated massive MISO CDI after successively transforming (translating and scaling) the source constellation following each feedback interval. 
This repositioning of the source constellation allows the feedback process to track the channel of each antenna. We assume that there is only a single Viterbi block at each user. Hence, unlike \cite{choi2013}, we do not minimize the paths over multiple blocks of parallel coherent decoders with different amplitude scalings and phase rotations, as this offers limited gain.
\vspace{-10pt}
\subsection{Transformed source constellation at each stage}
\begin{figure}[h]
\centering
\scalebox{.43}
{\includegraphics[width=20.5cm, height=16cm]{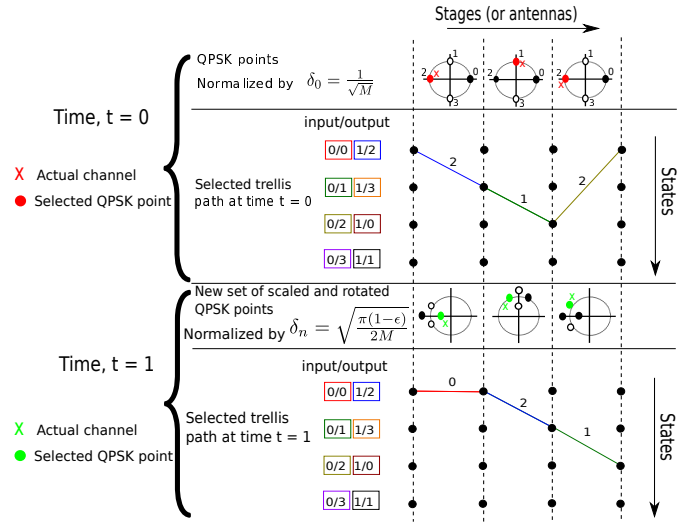}}\\
  \caption{An example of scaled and rotated QPSK constellation points at $t=1$ centered around the previously selected point at $t=0$ for up to three antenna channels.}
  \label{Fig0}
\end{figure}

The basic idea in the proposed scheme is to keep track of the selected source constellation points at each trellis stage and define a new  constellation for the next feedback centered around the previously selected constellation points. For the first feedback interval, we use the TCQ method described in Section \ref{4a}. Starting with the second feedback interval, the source constellation points at time $t$ are transformed for all the stages such that the previously selected source constellation points becomes the new centers of the transformed constellations at time $t+1$. For example, all four points in the original non-normalized QPSK constellation $[1,j,-1,-j]$ are transformed into new points using translation and scaling methods, to be discussed. Apart from this modification, the quantization process follows the TCQ approach discussed in Section \ref{4a}. An example\footnote{Due to the space constraint, we restrict to the example of QPSK constellation only.} of the proposed method with a QPSK constellation for the first 3 stages and $N=4$ is shown in Fig. \ref{Fig0}, where the first feedback at $t=0$ is illustrated at the top with red
dots representing the selected QPSK constellation points at each stage for the selected path $[2,1,2]$. At $t=1$, the transformed QPSK constellation at each stage is shown in the middle of Fig. \ref{Fig0} for the selected path $[0,2,1]$. At any given stage, the transformed QPSK constellation is centered around the previously selected QPSK point $(t=0)$, with a scaling factor $\delta_n$ (derived in Section \ref{ourb}). It should be stressed that this proposed transformation of the QPSK constellation is achieved at both BS and user without sharing any additional information through the feedback link. The BS also transforms the QPSK constellation after each feedback such that the new constellation at each stage is centered around the previously selected QPSK point.
\subsection{2D translation and scaling techniques}\label{ourb}
The TCQ method for massive MISO channels uses the Viterbi algorithm to
quantize the channel and unlike conventional MISO systems, it does not maintain a codebook that is scaled and rotated to the desired location. Hence, we introduce 2D translation and scaling transformations for the non-normalized source constellation at each stage in the trellis.
For example, with the non-normalized QPSK constellation $[1,j,-1,-j]$, in order to have the previously selected QPSK point, $\hat{x} [t-1]= \hat{a}[t-1]+ j \hat{b}[t-1]$ at the center of the transformed QPSK constellation, the translation of the $i^{\textrm{th}}$ QPSK point, $x_i= a+ jb$, along with scaling by $\delta_n$ is given by \cite{newman1979principles}
\begin{equation}\label{42}
\begin{bmatrix}
  \tilde{a}[t]  \\[0.3em]
  \tilde{b}[t]  \\[0.3em]
  1
  \end{bmatrix} = \begin{bmatrix}
       1 & 0 & \hat{a}[t-1]          \\[0.3em]
       0 & 1           & \hat{b}[t-1] \\[0.3em]
       0           & 0 & 1
     \end{bmatrix} \ \begin{bmatrix}
       \delta_n & 0 & 0          \\[0.3em]
       0 & \delta_n           & 0 \\[0.3em]
       0           & 0 & 1
     \end{bmatrix} \ \begin{bmatrix}
  a  \\[0.3em]
  b  \\[0.3em]
  1
  \end{bmatrix}, \\
\end{equation}
where $\tilde{x}_i [t]= \tilde{a}[t]+ j \tilde{b}[t]$ is the $i^{\textrm{th}}$ transformed QPSK point.
All points in the QPSK constellation are translated and scaled using \eqref{42}. Note that scaling and translation are performed on the non-normalized QPSK points.

In order to track the $m^{\textrm{th}}$ channel entry, $\bar{h}^{(m)}$, over time, the scaling factor, $\delta_n$, needs to be carefully designed, such that $\bar{h}^{(m)}$ lies close to the transformed source constellation points. We can define the mean channel variation due to the temporal correlation for the $m^{\textrm{th}}$ antenna as the mean Euclidean distance between the current and the previous normalized channel values, that is
\begin{align}
 \textrm{d}_{\textrm{mean}}&= \mathbb{E} \left[ \left| \bar{h}^{(m)}[t-1] -  \bar{h}^{(m)}[t]\right| \right],\\
 & = \mathbb{E} \left[ \left| \frac{h^{(m)}[t-1]}{\|\mathbf{h}[t-1] \|} - \frac{ h^{(m)}[t]}{\|\mathbf{h}[t] \|} \right| \right],\label{vr}
\end{align}
where $h^{(m)}[t-1]$ and $h^{(m)}[t]$ denote the $m^{\textrm{th}}$ entry of the channel vectors $\mathbf{h}[t-1]$ and $\mathbf{h}[t]$, respectively. Due to channel hardening caused by the large dimensions of $\mathbf{h}[t]$ and $\mathbf{h}[t-1]$, $\frac{\|\mathbf{h}[t] \|^2}{M} $ and $\frac{\|\mathbf{h}[t-1] \|^2}{M} $ approach 1, and thus
\begin{align}
 \frac{h^{(m)}[t]}{\|\mathbf{h}[t] \|} & = \frac{h^{(m)}[t]/\sqrt{M}}{\sqrt{\|\mathbf{h}[t] \|^2 /M}} \approx \frac{h^{(m)}[t]}{\sqrt{M}}.
\end{align}
This allows us to approximate \eqref{vr} as
\begin{align}
 &\textrm{d}_{\textrm{mean}} \approx \frac{1}{\sqrt{M}}\mathbb{E} \left[ \left| h^{(m)}[t-1] -  h^{(m)} [t] \right| \right]\nonumber \\
 & = \frac{1}{\sqrt{M}} \mathbb{E} \left[ \left| h^{(m)}[t-1] - \left( \epsilon h^{(m)}[t-1]+ \sqrt{1-\epsilon^2} g^{(m)}[t]  \right) \right| \right] \nonumber \\
 & = \frac{1}{\sqrt{M}} \mathbb{E} \left[ \left| h^{(m)}[t-1] \left(1- \epsilon \right) - \sqrt{1-\epsilon^2} g^{(m)}[t]   \right| \right], \label{13}
\end{align}
where $g^{(m)}[t] $ denotes the $m^{\textrm{th}}$ entry of the vector $\mathbf{g}[t]$, (the same vector as $\mathbf{g}_k [t]$ in \eqref{0}). Denoting $\Re \{h^{(m)}[t-1]\}$ and $ \Im \{h^{(m)}[t-1]\}$, $\Re \{g^{(m)}[t]\}$ and $\Im \{g^{(m)}[t]\}$ as the real and imaginary parts of $h^{(m)}[t-1]$ and $g^{(m)}[t]$, respectively, we have
\begin{align}
 \textrm{d}_{\textrm{mean}} \approx \frac{1}{\sqrt{M}} \mathbb{E} \left[ \sqrt{ X^2 + Y^2}\right],
  \label{14}
 \end{align}
where
\begin{equation}
X= \left(1- \epsilon \right) \Re \{h^{(m)}[t-1]\} - \sqrt{1-\epsilon^2} \Re \{g^{(m)}[t]\}, \nonumber
 \end{equation}
and
\begin{equation}
Y= \left(1- \epsilon \right) \Im \{h^{(m)}[t-1]\} -\sqrt{1-\epsilon^2} \Im \{g^{(m)}[t]\}.
\end{equation}
Since $\Re \{h^{(m)}[t-1]\}$, $\Re \{g^{(m)}[t]\}$, $\Im \{h^{(m)}[t-1]\}$ and $\Im \{g^{(m)}[t]\}$ are independent and distributed according to the normal distribution $\mathcal{N}(0,1/2)$, the random variables $X$ and $Y$ are $\mathcal{N}(0,1-\epsilon)$. 
Therefore, we can define $Z=\sqrt{X^2 + Y^2}$ to be Rayleigh distributed with scale parameter $\sigma = \sqrt{1-\epsilon}$,
and the mean value $\sqrt{1-\epsilon}\sqrt{\pi/2}$ \cite{kotz1994continuous}. We can thus rewrite \eqref{14} as
\begin{equation}\label{dm}
 \textrm{d}_{\textrm{mean}} \approx \sqrt{\frac{\pi(1-\epsilon)}{2M}}.
\end{equation}
The mean channel variation includes the effects of both temporal correlation and the number of transmit antennas at the base station. It is noted that channel entries change slowly over time as $M$ increases. In order to track the slow varying channels  and to have the source constellation points closer to each other for fine quantization, we use \eqref{dm} as the scaling value, $\delta_n$, for the source constellation, after the first feedback interval, such that
\begin{align}
  \delta_n =  \sqrt{\frac{\pi(1-\epsilon)}{2M}},  \ \ \ \ t > 0.
  \label{50}
\end{align}
The initial scaling factor for the source constellation, $\delta_0 = 1/\sqrt{M}$, is used only for the first feedback interval. Following the first feedback interval, the  value used to scale the source constellation for the remaining feedback intervals is $\delta_n$. Both BS and user compute the scaling value in \eqref{50} using the temporal correlation coefficient and the number of BS antennas. We assume that perfect knowledge of $\epsilon$ is available at the BS and the user. The scaling parameter in \eqref{50} needs not to be computed after each feedback interval as long as the temporal correlation statistic of the channel remains the same.

The proposed differential TCQ method for the temporally correlated MISO channel does not increase the feedback overhead and performance improvements are achieved via the systematic translation and scaling of the source constellation.
It is important to note that the NTCQ differential scheme \cite{choi2013} depends on two crucial operations: a) finding the null space of the temporally correlated channel vector before quantization and b) selecting the appropriate weights after quantization. In contrast, the proposed TCQ scheme relies solely on the transformation of the source constellation and does not require finding optimal weights, thus reducing the complexity. The feedback overhead is also reduced in the proposed scheme, as it does not have to feedback additional bits (for optimal weights) to the BS. 
\vspace{-5pt}
\section{Proposed TCQ for Spatially correlated channels}\label{our_sp}
In this section, we present a TCQ scheme to quantize spatially correlated channels. In conventional MIMO systems, the codebook design for a  spatially correlated channel is fixed where codebook entries are directed towards dominant eigenvectors of the transmit correlation matrix \cite{raghavan2007systematic}. For massive MISO systems, codebook based limited feedback techniques are not practically feasible. Therefore an alternate method is needed to quantize a spatially correlated channel. For this purpose, an adaptive TCQ based limited feedback method is developed in \cite{choi2013} that decorrelates the channel at the user before quantization. The drawback of this method is that it requires the knowledge of spatial correlation matrix at the transmitter. In this paper, the spatially correlated massive MISO channel is modeled by 
\begin{equation}\label{sp_corr}
 \mathbf{h} = \mathbf{h}_{\textrm{iid}} \mathbf{R}^{1/2},
\end{equation}
where $\mathbf{h}_{\textrm{iid}}$ is a $1 \times M$ vector with entries distributed according to $\mathcal{CN} (0,1)$. $\mathbf{R}$ is the transmit spatial correlation matrix. In this paper, we assume that entries of $\mathbf{R}$, $r_{ij}$, follow an exponential correlation model \cite{951380,choi2014}
\begin{equation}
r_{ij} = z_t^{d_{ij}},
\end{equation}
where $d_{ij}$ is the distance between the antenna $i$ and $j$. $z_t$ is a spatial correlation coefficient with $0 \leq z_t \leq 1$, where 0 represents no spatial correlation and 1 represents a fully correlated channel. We consider ULA and URA antenna topologies at the BS. The physical dimension of the one-dimensional ULA is much larger than that of URA, therefore the latter is practically more suitable for massive MISO deployments. We constrain the dimensions of the URA, such that for a given value of $M$, the dimensions of the URA are $\sqrt{M} \times \sqrt{M}$.

\begin{figure}[!t]
\vspace{-1em}
   \centering
\includegraphics[width=8.8cm, height=6.5cm]{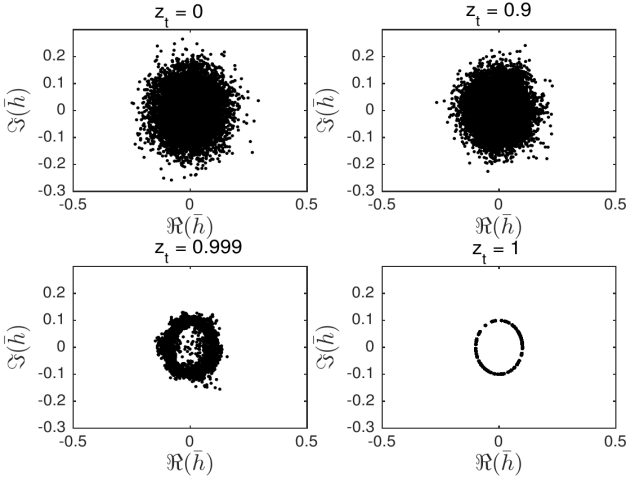}\\
  \caption{Distribution of channel entries of the normalized channel for $M=100$ and different values of $z_t$.}
  \label{Figsp}
  \vspace{-5pt}
\end{figure}

In order to design a TCQ scheme for spatially correlated massive MISO channels, we first analyze the impact of $z_t$ on the distribution of channel entries. For this purpose, we plot the channel entries of the normalized spatially correlated channel with $M=100$, over a large number of channel realizations in Fig. \ref{Figsp}, for various values of $z_t$. It is important to note that Fig. \ref{Figsp} applies to both ULA and URA topologies. 
If $\bar{{h}}^{(i)}$ is the $i^{\textrm{th}}$ element of $\mathbf{h}/\| \mathbf{h}\|$ and $h^{(i)}$ is the $i^{\textrm{th}}$ element of $\mathbf{h}$, then in the i.i.d. case ($z_t = 0$), $|\bar{h}^{(i)}|^2 = |{h}^{(i)}|^2/ \sum_{j=1}^{M} |{h}^{(j)}|^2$ and since each $|{h}^{(j)}|^2$ is a standard exponential, $|\bar{h}^{(i)}|^2$ has a Beta distribution. Hence, the amplitude of $\bar{h}^{(i)}$ is the square root of the a Beta variable with range, $0 < |\bar{h}^{(i)}| < 1$. Since the phase of ${h}^{(i)}$ is uniform on $[0, 2\pi]$, it follows that the normalized channel entries are isotropic, as shown by the circular pattern in Fig. \ref{Figsp}. At the opposite extreme, when $z_t = 1$, we have perfectly or fully correlated entries. Denoting the $i^{\textrm{th}}$ element of the fully correlated channel by $\bar{h}^{(i)}_f$, where $\bar{h}^{(i)}_f = \bar{h}_f$ $\forall i$, then $|\bar{h}^{(i)}_f|^2 = |{h}^{(i)}_f|^2/ \sum_{j=1}^{M} |{h}^{(j)}_f|^2 = 1/M$. Hence, for $z_t = 1$, the channel entries lie on a circle of radius $1/\sqrt{M}$, as shown in Fig. \ref{Figsp}. Clearly as $z_t$ varies from 0 to 1, the channel entry distribution changes from a circular spread with random radius to a fixed circle of radius $1/\sqrt{M}$.
 
From Fig. \ref{Figsp}, we also make the
following observation. As, the value of $z_t$ increases, the distribution of entries of the normalized spatially correlated channel shows less spread compared to i.i.d. Rayleigh channels ($z_t=0$). However, this trend is mainly visible for high values of $z_t$, such that $z_t \to 1$. Even with $z_t=0.9$, the spread of the normalized channel entries is similar to the $z_t=0$ case.   

Motivated by these observations, we propose a TCQ scheme for highly spatially correlated MISO channels, where we select the scaling parameter, $\hat{\delta_n}$, for the source constellation that gives the minimum mean squared error (MMSE) between the entries of the fully correlated normalized channel ($z_t = 1$) and the scaled source constellation, given by
\begin{equation} \label{mmse}
 \hat{\delta}_n = \textrm{arg} \min_{\substack{0 \leq i \leq L \\ \hat{\delta_n} \geq 0}} \mathbb{E} \left[ |\bar{{h}}_f- \hat{\delta}_n x_i |^2\right],
\end{equation}
where $x_i$ corresponds to the $i^{\textrm{th}}$ point in the source constellation. The solution that minimizes the MMSE in \eqref{mmse} is equal to the standard deviation of the channel entries, $\bar{{h}}_f$, given by $1/\sqrt{M}$. Therefore, for the proposed method, we set the scaling parameter for the source constellation as
\begin{equation}
 \hat{\delta}_n = \frac{1}{\sqrt{M}}.
\end{equation}

\begin{figure}[!t]
    \centering
        \scalebox{.59}
     {\includegraphics[width=8.5cm, height=4.4cm]{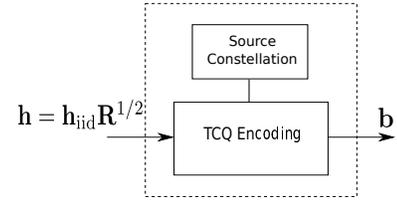}}
      \caption{The block diagram of channel quantization at the user for spatially correlated channels.}\label{fig:sp}
    \end{figure}

The proposed method uses the TCQ method discussed in Section \ref{4a} to quantize the channel, $\mathbf{h}$, in \eqref{sp_corr}, where the source constellation (either QPSK or 8PSK) used is scaled by $\hat{\delta}_n$ and is fixed across all the stages in the trellis. 
The block diagram of the proposed TCQ technique at the user/receiver is shown in Fig. \ref{fig:sp}. Note that as discussed in Section \ref{4a} and \ref{ourr}, the channel vector $\mathbf{h}$ is normalized before TCQ encoding.
The channel reconstruction at the transmitter follows the same procedure discussed in Section \ref{4a}.  
    
Unlike \cite{choi2013}, the proposed method does not decorrelate the channel at the user in order to quantize $\mathbf{h}_{\textrm{iid}}$, instead it directly quantizes the channel $\mathbf{h}$. Therefore, the proposed method, unlike \cite{choi2013}, does not require the knowledge of the correlation matrix at the transmitter to reconstruct the correlated channel, hence reducing the feedback overhead of additional information to the transmitter. The performance of the proposed TCQ scheme for spatially correlated channels is demonstrated in Section \ref{num} with both ULA and URA topologies.   
\vspace{-6pt}
\section{Beamforming gain}
The performance of the TCQ based limited feedback techniques proposed for temporally and spatially correlated channels presented in Section \ref{ourr} and Section \ref{our_sp}, respectively, can also be evaluated in terms of the average beamforming gain.  
The beamforming gain of the MISO system at time $t$, for any single user, is defined as \cite{choi2013}
\begin{equation}
 \textrm{BF} = \left|\mathbf{h}[t] \tilde{\mathbf{h}}^H[t]\right|^2,
\end{equation}
where the beamforming vector is denoted by $\tilde{\mathbf{h}}$, which is the quantized version of the CDI, such that $\|\tilde{\mathbf{h}}\| = 1$. The average beamforming gain can be expressed as
\begin{align}
 \mathbb{E} \left[\textrm{BF}\right]  = \mathbb{E} \left[\left|\mathbf{h}[t] \tilde{\mathbf{h}}^H[t]\right|^2\right] 
  = M \mathbb{E} \left[\left|\bar{\mathbf{h}}[t] \tilde{\mathbf{h}}^H[t]\right|^2\right]. \label{bf_gain}
\end{align}
The average beamforming gain in \eqref{bf_gain} is evaluated numerically and used as a performance metric in Section \ref{num}. 
\vspace{-10pt}
\section{Numerical Results}\label{num}
In this section, we present simulation results for the proposed differential TCQ scheme and the proposed spatially correlated TCQ scheme in massive MISO systems. We compare the performance of the proposed schemes with the differential NTCQ and the adaptive NTCQ techniques in \cite{choi2013}. In order to evaluate the performance of the proposed schemes, we consider three types of channels: the temporally correlated channel \eqref{0}, the spatially correlated channel \eqref{sp_corr} and the standardized WINNER II channel. 
\begin{figure}[!b]
\centering
\vspace{-15pt}
\includegraphics[width=7.5cm, height=5.2cm]{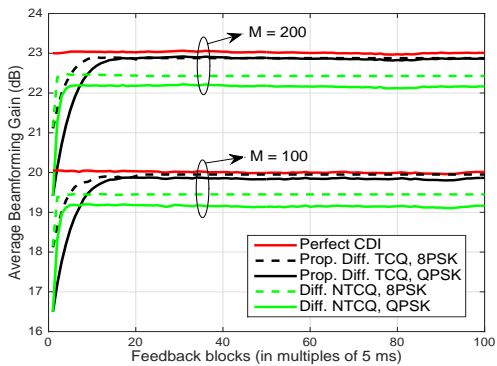}
  \caption{Average beamforming gain versus feedback intervals with $M=100$ and $M=200$ for $\epsilon = 0.9881 (v= 3$ km/h$)$.}
  \label{Fig3}
\end{figure}
In the case of a MU MISO system we assume a constant ratio $q=M/K=10$, unless stated otherwise. The temporal correlation coefficient, $\epsilon$, follows Jake's model, such that $\epsilon = J_0 (2\pi f_d T) $, where $J_0$ is the zeroth order Bessel function of the first kind, $f_d$ is a Doppler frequency and $T$ is the channel feedback interval. In this paper, the feedback interval is set to $T = 5$ ms which is typically used to evaluate the performance of limited feedback systems. Generally, the feedback interval is set less than the coherence time of the channel, so that the BS gets adequate time to compute beamforming vectors for the users. The carrier frequency is $2.5$ GHz. Although, in practical settings, the feedback links are lossy with delays, in order to focus the study on the performance of the CDI quantization process, we assume that the feedback link is lossless with zero delay in the simulations. For a fair comparison between the proposed schemes and the NTCQ schemes  \cite{choi2013}, the source constellations used are QPSK and 8PSK with $N=4$ and $N=8$, respectively. A single block of the trellis decoder is considered at the receiver.

In the figures, we refer to the differential NTCQ method of \cite{choi2013} for temporally
correlated channels as ``Diff. NTCQ''. The proposed differential TCQ method is referred to as ``Prop. Diff. TCQ''. Similarly, for spatially correlated channels, we refer to the proposed TCQ scheme as ``Prop. spatial corr.'' and the adaptive NTCQ scheme in \cite{choi2013} is referred as ``Adaptive spatial corr.''.
\vspace{-10pt}
\subsection{Temporally correlated channels}
For temporally correlated channels, we use the channel model in \eqref{0} and present the performance of the proposed differential TCQ scheme in MU massive MISO systems.
\begin{figure}[!b]
\vspace{-15pt}
  \centering
  \subfigure[QPSK]{%
    \includegraphics[width=7.5cm, height=5.2cm]{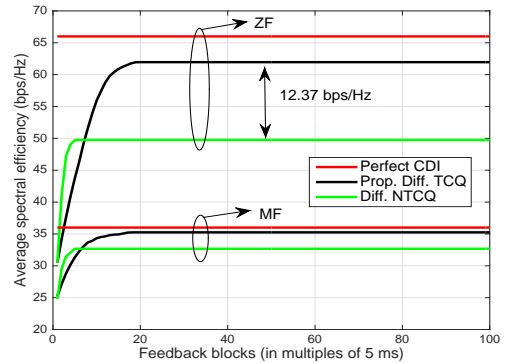}%
    \label{fig:x1}%
  }\\
\vspace{-10pt}
   \subfigure[8PSK]{%
    \includegraphics[width=7.5cm, height=5.2cm]{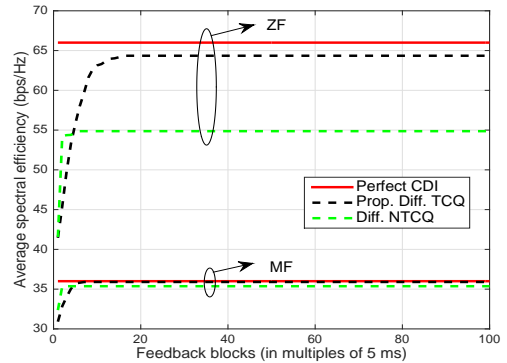}%
    \label{fig:x2}%
    }
  \caption{Average spectral efficiency versus time with $M=100$ and $\epsilon = 0.9881$ ($v= 3$ km/h).}
  \label{xxxx}
\end{figure}
\subsubsection{Beamforming gain}
We use the average beamforming gain metric \cite{choi2013} defined in \eqref{bf_gain} to evaluate the performance of a MISO system. Fig. \ref{Fig3} shows average beamforming gain results against 100 feedback blocks (size of each block is 5 ms) for the user velocity $v = 3$ km/h ($\epsilon$ = 0.9881) with $M=100$ and $M=200$. The proposed differential scheme provides approximately 1 dB gain compared to  \cite{choi2013} for both $M=100$ and $M=200$ with a QPSK constellation. The average beamforming gain with the 8PSK constellation is higher than the QPSK constellation because the former uses 2 bits to quantize each antenna channel, resulting in smaller quantization errors.

It is important to note that the proposed scheme requires $M$ and $2M$ feedback bits with QPSK and 8PSK constellations, respectively, whereas an additional 4 bits are required in the differential NTCQ method \cite{choi2013} to equip the BS with optimal weights, increasing the total
number of feedback bits to $M+4$ and $2M+4$ for QPSK and 8PSK, respectively. Thus, in addition to the improved beamforming gain, the proposed method also reduces the feedback overhead.
%
%
%
\subsubsection{MU spectral efficiency and SINR performance}
\begin{figure}[!t]
  \centering
  \subfigure[QPSK]{%
    \includegraphics[width=7.5cm, height=5.2cm]{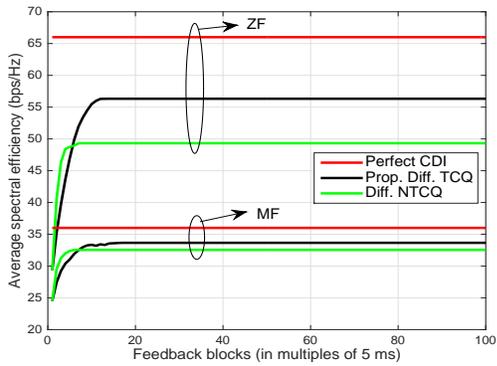}%
    \label{fig:x11}%
  }\\
\vspace{-10pt}
   \subfigure[8PSK]{%
    \includegraphics[width=7.5cm, height=5.2cm]{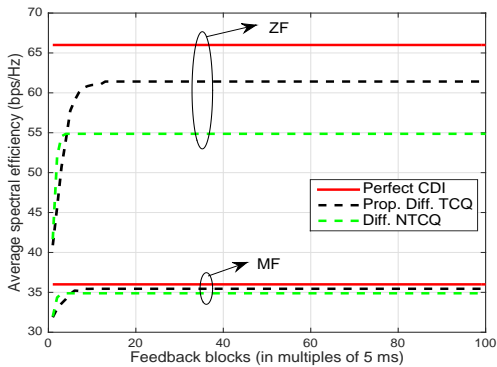}%
    \label{fig:x22}%
    }
  \caption{Average spectral efficiency versus time with $M=100$ and $\epsilon = 0.9672$ ($v= 5$ km/h).}
  \label{xxxx1}
\end{figure}
\begin{figure}[!t]
  \centering

  \subfigure[QPSK]{%
    \includegraphics[width=7.5cm, height=5.2cm]{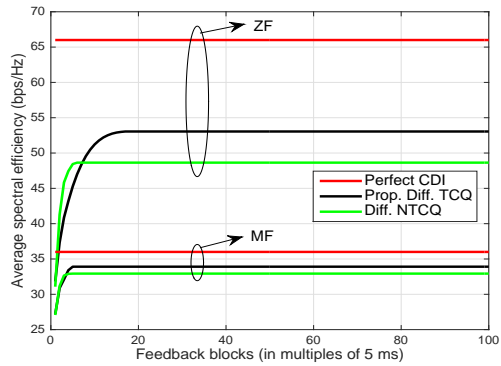}%
    \label{fig:x111}%
  }\\
\vspace{-10pt}
   \subfigure[8PSK]{%
    \includegraphics[width=7.5cm, height=5.2cm]{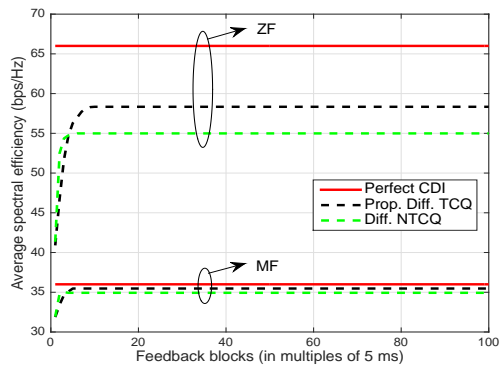}%
    \label{fig:x222}%
    }
  \caption{Average spectral efficiency versus time with $M=100$ and $\epsilon = 0.9363$ ($v= 7$ km/h).}
  \label{xxxx2}
\end{figure}

For MU MISO systems, we use ZF precoding and MF precoding to compute average SINR and spectral efficiency results. Figs. \ref{fig:x1} and \ref{fig:x2} show the average spectral efficiency versus feedback interval for a MU massive MISO system at $\rho= 10$ dB, with $M=100$ and $v=3$ km/h ($\epsilon = 0.9881$) with QPSK and 8PSK constellations, respectively. ZF precoding outperforms MF precoding in the temporally correlated channels. We also note that the spectral efficiency performance of the proposed scheme exceeds that of the differential NTCQ method. Although, the NTCQ scheme takes less feedback intervals to reach the maximum beamforming gain, the proposed scheme starts to yield better performance as the number of feedback intervals increases. In Fig.~\ref{fig:x1}, the proposed differential TCQ scheme provides nearly 13 bps/Hz and 2 bps/Hz average spectral efficiency gains compared to the differential NTCQ method with QPSK constellation, for ZF and MF precoding schemes, respectively.

A similar trend is seen in Fig. \ref{fig:x2} with 8PSK constellation for ZF precoding, but for MF precoding, the performance gap between the proposed differential TCQ scheme and the NTCQ scheme decreases. As the velocity increases to $v=5$ km/h, the variations in the channel also increase which leads to a higher spectral efficiency loss with ZF precoding, as seen in Fig. \ref{fig:x11} and Fig. \ref{fig:x22}. However, the spectral efficiency results with MF precoding remain stable as velocity increases. A similar trend is seen in Fig. \ref{fig:x111} and Fig. \ref{fig:x222}, for $v=7$ km/h. The proposed differential TCQ scheme exhibits higher spectral efficiency loss at high velocities compared to the differential NTCQ scheme with ZF precoding, but still performs better than the differential NTCQ scheme. Table \ref{Tab:SRNRValues} shows the average spectral efficiency of the proposed differential TCQ and the differential NTCQ scheme for various speed values with ZF precoding and QPSK constellation. The proposed differential TCQ yields better average spectral efficiency for slow speeds compared to the differential NTCQ scheme. However, at $v=13$ km/h, both schemes have the same performance.
\begin{table}[!b]
\caption{Average spectral efficiency of the proposed differential TCQ scheme and the differential NTCQ scheme.}
\label{Tab:SRNRValues}
\begin{center}
\resizebox{220pt}{26pt}{
\begin{tabular}{|c|c|c|c|c|}

\hline
  & \multicolumn{1}{c|}{$v=3$ km/h}  & \multicolumn{1}{c|}{$v=7$ km/h} &  \multicolumn{1}{c|}{$v=13$ km/h} \bigstrut \\
   & \multicolumn{1}{c|}{$\epsilon=0.9881$} &  \multicolumn{1}{c|}{$\epsilon=0.9363$ } &  \multicolumn{1}{c|}{$\epsilon=0.7895$} \bigstrut \\ \hline
Prop. Diff. TCQ & 60 bps/Hz&  52 bps/Hz& 39 bps/Hz\\
\hline
Diff. NTCQ & 50 bps/Hz& 47 bps/Hz&  39 bps/Hz \\
\hline
\end{tabular}
}
\end{center}
\end{table}
 \begin{figure}[!t]
   \centering
  \includegraphics[width=7.5cm, height=5.2cm]{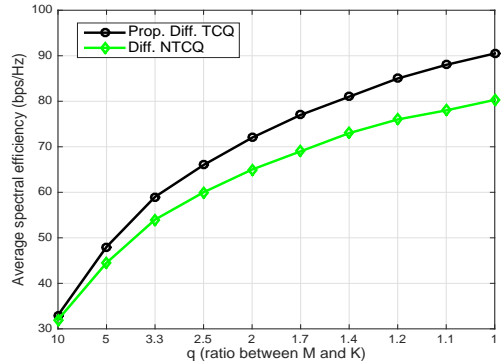}
  \caption{Average spectral efficiency for various $q=M/K$ values with QPSK constellation, $M=100$ and $\epsilon = 0.9881 \ (v= 3$ km/h$)$.}
  \label{Fig11_1}
\end{figure}
%
%
 \begin{figure}[!t]
   \centering
  \includegraphics[width=7.5cm, height=5.2cm]{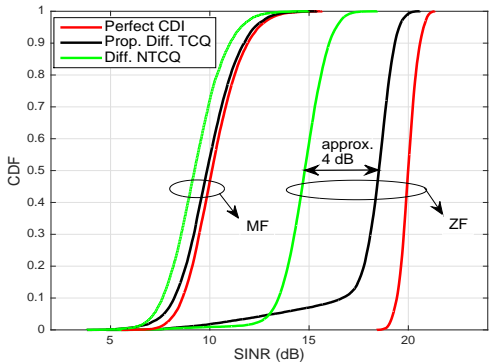}
  \caption{CDF of the SINR for an arbitrary user with QPSK constellation, $M=100$ and $\epsilon = 0.9881$ ($v= 3$ km/h).}
  \label{Fig7}
\end{figure}


Figure \ref{Fig11_1} shows the average spectral efficiency against different values of $q$ for the proposed differential TCQ and differential NTCQ schemes using a QPSK constellation with $M=100$, $v=3$ km/h and $\rho=10$ dB. The proposed scheme achieves better spectral efficiency than the differential NTCQ scheme, as the former yields less quantization error. It is seen that as more users are added in the system, the performance gain with the proposed scheme increases compared to the differential NTCQ scheme. 

The per user SINR cumulative distribution function (CDF) is shown in Fig. \ref{Fig7} at $\rho=10$ dB with a QPSK constellation. For ZF precoding, the SINR CDF of the proposed differential TCQ scheme has a long-tail. This is because the basic TCQ method (discussed in Section \ref{4a}) is used for the first feedback yielding low SINR performance, but with time the SINR performance improves using the proposed differential TCQ method. The SINR CDFs confirm the spectral efficiency results i.e., the mean SINR of the proposed differential TCQ scheme is greater than the differential NTCQ method for both ZF and MF precoding schemes. For example by looking at the median values of SINR CDFs with ZF precoding in Fig \ref{Fig7}, the difference between median SINR values for the proposed scheme, $\textrm{SINR}_{\textrm{prop}}$, and the NTCQ scheme, $\textrm{SINR}_{\textrm{NTCQ}}$, is $4 \ [\textrm{dB}]$ or on a linear scale, $\textrm{SINR}_{\textrm{prop}} = 2.5 \times \textrm{SINR}_{\textrm{NTCQ}}$, which gives a spectral efficiency gain relative to differential NTCQ scheme as $K \log_2 (1 + \textrm{SINR}_{\textrm{prop}}) - K \log_2 (1 + \frac{\textrm{SINR}_{\textrm{prop}}}{2.5}) $. In Fig \ref{Fig7}, for ZF precoding, the median value of $\textrm{SINR}_{\textrm{prop}}$ is $18.5$ dB and $\textrm{SINR}_{\textrm{NTCQ}}$ is $14.5$ dB, therefore, the gain in the spectral efficiency relative to the differential NTCQ scheme is $12.37$ bps/Hz, which is also evident in Fig.~\ref{fig:x1}.

\vspace{-10pt}
\subsection{Spatially correlated channels}
For spatially correlated channels, we use the channel model in \eqref{sp_corr} and present the performance of the proposed spatially correlated TCQ scheme (discussed in Section \ref{our_sp}) for MU massive MISO systems. We consider both ULA and URA antenna settings, as discussed in Section \ref{our_sp}.
\begin{figure}[!h]
   \centering
  \includegraphics[width=7.5cm, height=5.2cm]{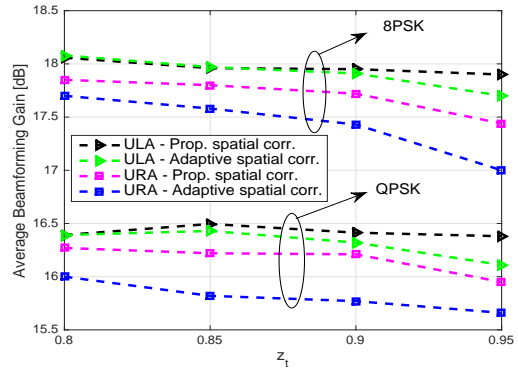}
  \caption{Average beamforming gain for different values of $z_t$ with  $M=100$.}
  \label{Fig8}
\end{figure}
 \begin{figure}[!h]
   \centering
  \includegraphics[width=7.5cm, height=5.2cm]{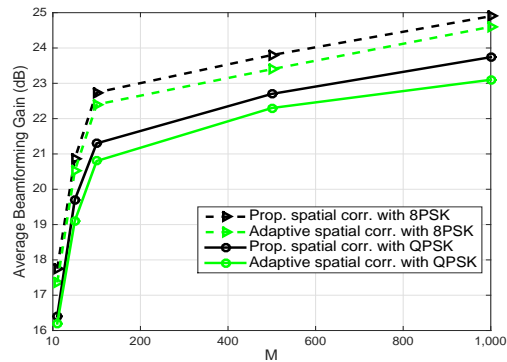}
  \caption{Average beamforming gain for different values of $M$ with $z_t = 0.99$ and ULA.}
  \label{Fig9}
\end{figure}

\subsubsection{Beamforming gain} In Fig. \ref{Fig8}, we compare the proposed spatially correlated TCQ scheme and the adaptive TCQ technique \cite{choi2013}, in high spatially correlated channels, such that $z_t \to 1$. The comparison is performed in terms of the average beamforming gain for QPSK and 8PSK constellations with ULA and URA. It is seen in Fig. \ref{Fig8} that the proposed spatially correlated TCQ scheme performs slightly better than the adaptive TCQ with both QPSK and 8PSK constellations. 

It is important to mention that the proposed scheme does not require any knowledge (perfect or partial) of the transmit correlation matrix, $\mathbf{R}$, at the transmitter. On the other hand, it is assumed in \cite{choi2013} that perfect knowledge of $\mathbf{R}$ is available at the transmitter. Unlike \cite{choi2013}, the proposed spatially correlated TCQ scheme quantizes the channel without decorrelating the spatially correlated channel. It is also noted that the ULA provides higher beamforming gain compared to the URA. This is due to the fact that the URA induces higher spatial correlation in the channel than the ULA, thus resulting in reduced beamforming gain. 
We plot average beamforming gain in Fig. \ref{Fig9} for different values of $M$ with fixed $z_t = 0.99$. At higher values of $M$, the gain of the proposed spatially correlated TCQ scheme is approximately 0.5 dB and 1 dB higher than the adaptive TCQ scheme, with QPSK and 8PSK constellations, respectively.

\subsubsection{MU spectral efficiency performance} 
\begin{figure}[!t]
   \centering
  \includegraphics[width=7.5cm, height=5.2cm]{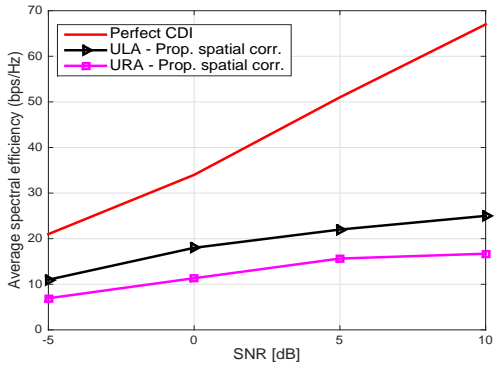}
  \caption{Average spectral efficiency performance against different SNR values for $M=100$, $q=10$ and $z_t = 0.9$.}
  \label{Fig10}
\end{figure}

The average spectral efficiency of the MU MISO system is shown in Fig. \ref{Fig10} for the proposed spatially correlated TCQ scheme with $M=100$, $z_t = 0.9$ using the 8PSK constellation. ZF precoding is used to compute precoding vectors at the BS. It is noticed that the spatial correlation reduces the spectral efficiency of the TCQ based limited feedback system. Also, interference due to the quantization process is a limiting factor that causes the spectral efficiency to saturate at high SNR values. Due to higher numbers of adjacent antennas, the correlation in the URA is higher compared to the ULA, which results in a reduced spatial separation between users. Therefore, we note that the ULA provides significant improvement in the spectral efficiency compared to the URA. The dominance of ULA compared to other antenna topologies, in terms of the spectral efficiency, has also been noted in \cite{6736761,1285039,abouda2006effect,tufvesson_mimo}.

\vspace{-10pt}
\subsection{WINNER II channels}
\begin{figure}[!t]
   \centering
  \includegraphics[width=7.5cm, height=5.2cm]{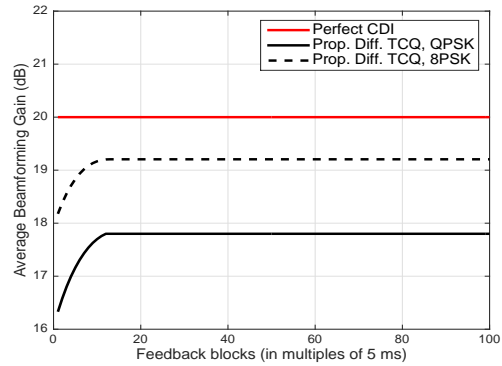}
  \caption{Average beamforming gain versus feedback intervals with $M=100$ for $\epsilon = 0.9987 \ (v= 1$ km/h$)$.}
  \label{Fig11}
\end{figure}

We also evaluate the performance of the proposed differential TCQ scheme in the standardized WINNER channel model \cite{ist2007deliverable}. The WINNER II channel model is based on IMT-Advanced (M.2135) channel model recommended by ITU-R. For the WINNER II channel, we use $M=100$ transmit antennas in a ULA setting with $0.5 \lambda$ spacing between them, where $\lambda$ denotes the wavelength of the carrier frequency. We consider an urban macro (UMa) scenario with non-line-of-sight (NLoS) propagation. The carrier frequency is $2.5$ GHz and the velocity of the receiver is $v = 1$ km/h. 
For the parameters discussed above, we generate a WINNER II MISO channel and plot the average beamforming gain for the massive MISO system with both QPSK and 8PSK constellations. As the WINNER II channel model is both temporally and spatially correlated, we rely on the differential TCQ scheme (discussed in Section \ref{4a}) to keep track of the channel entries over time. Assuming that the velocity and Doppler frequency is perfectly estimated at the receiver, we compute corresponding temporal coefficient, $\epsilon$, as defined in FOGM channels, in order to evaluate the scaling parameter for WINNER II channels. The result shown in Fig. \ref{Fig11} has a similar trend to that of the temporally correlated channel generated with the first-order Gauss-Markov process. This validates the performance of the proposed differential TCQ scheme in real-world channels.
\vspace{-8pt}
\section{Conclusion}
One of the major advantages of the TCQ technique is that the selection and searching of the appropriate codeword becomes simpler using the Viterbi algorithm. For MU transmission, we have shown that RVQ codebooks require very large numbers of codewords in a massive MISO system that increases the search complexity for the appropriate codeword enormously. In doing so, we derived the expected SINR approximation for the MF precoding scheme with RVQ CDI, and used the expected SINR approximation to compute the number of bits required to achieve an expected SINR performance with RVQ codebooks which is $z$ dB below the expected SINR with perfect CDI.

In this paper, we have proposed efficient TCQ schemes to quantize temporally correlated channels in a massive MISO FDD systems. In the proposed differential TCQ scheme, we transform the source constellation at each stage in a trellis separately, such that the resulting constellation is centered around the previously selected constellation point at that particular stage. Consequently, we have introduced 2D translation and scaling techniques to transform the source constellation.
We have derived the scaling factor that exploits the temporal correlation present in the channel and scales the constellation accordingly for the given number of BS antennas and the temporal correlation coefficient.
For spatially correlated channels, we propose a method where the correlated channel is quantized using TCQ approach. The advantage of quantizing the correlated channel directly without using decorrelation at the receiver is that the transmitter can reconstruct the quantized channel without requiring the knowledge of the spatial correlation matrix.
Finally, we have shown via simulations that proposed TCQ methods outperform the existing NTCQ methods for both temporally and spatially correlated channels, by improving the spectral efficiency and beamforming gain of the system and reducing the feedback overhead.

The proposed schemes can readily be extended to MIMO channels. One possible solution is to quantize each received antenna channel vector using TCQ approach. The extension to massive MIMO will linearly increase the feedback overhead by the amount equal to the number of receive antennas. The TCQ schemes are also applicable to the single stream massive MIMO transmission, however, some
details of the procedure would need to be adjusted in this scenario. This is left for future 
work
\vspace{-12pt}
\appendix
\section{Expected SINR for MF precoding with perfect CDI}\label{B}
The expectation of the SINR in \eqref{111} can be written as
\begingroup\makeatletter\def\f@size{9}\check@mathfonts\def\maketag@@@#1{\hbox{\m@th\normalsize\normalfont#1}}
 \begin{equation}\label{exsinr}
 \mathbb{E} \left[\textrm{SINR}_k\right] = \mathbb{E} \left[\frac{ \frac{\rho}{K} \left| \mathbf{h}_k \mathbf{w}_k \right|^2 }
 {\frac{\rho}{K} \sum_{j \neq k}^{K} \left| \mathbf{h}_k \mathbf{w}_j \right|^2 +1}\right] = \mathbb{E}\left[ \frac{A}{B}\right].
\end{equation}\endgroup
Using the approach given in \cite{5510182} to evaluate the expectation over numerator, $A$, and denominator, $B$, where both are functions of the same random variables, and $M \to \infty$, we can write \eqref{exsinr} as
\vspace{-5pt}
\begingroup\makeatletter\def\f@size{9}\check@mathfonts\def\maketag@@@#1{\hbox{\m@th\normalsize\normalfont#1}}
\begin{equation}\label{exsinrreal}
 \mathbb{E} \left[\textrm{SINR}_k\right] \approx \frac{ \frac{\rho}{K} \mathbb{E} \left[\left| \mathbf{h}_k \mathbf{w}_k \right|^2\right] }
 {\frac{\rho}{K} \sum_{j \neq k}^{K} \mathbb{E} \left[\left| \mathbf{h}_k \mathbf{w}_j \right|^2\right] +1}.
 \end{equation}\endgroup
 
For MF precoding with perfect CDI, we have $\mathbf{w}_k = \bar{\mathbf{h}}_k^H$ and the expected SINR approximation for the $k^{\textrm{th}}$ user becomes
\begingroup\makeatletter\def\f@size{9}\check@mathfonts\def\maketag@@@#1{\hbox{\m@th\normalsize\normalfont#1}}
\begin{align}
\mathbb{E}\left[\textrm{SINR}_k^{\textrm{MF}}\right] & 
\approx \frac{ \frac{\rho}{K} \mathbb{E} \left[ \| \mathbf{h}_k \|^2 \right] \mathbb{E} \left[ | \bar{\mathbf{h}}_k \bar{\mathbf{h}}_k^H |^2 \right]}
{\frac{\rho}{K} \mathbb{E} \left[ \| \mathbf{h}_k \|^2 \right]  \sum_{j \neq k}^{K} \mathbb{E} \left[| \bar{\mathbf{h}}_k \bar{\mathbf{h}}_j^H |^2 \right]+1},\label{222}
\end{align}\endgroup
where we use the fact that amplitude and direction of $\mathbf{h}_k$ are independent.
In \eqref{222}, $\mathbb{E} [\| \mathbf{h}_k \|^2] = M$ and $\mathbb{E}[| \bar{\mathbf{h}}_k \bar{\mathbf{h}}_k^H |^2] =1 $. Also, the quantity $ \mathbb{E} [| \bar{\mathbf{h}}_k \bar{\mathbf{h}}_j^H |^2] = \frac{1}{M}$, as $\bar{\mathbf{h}}_k$ and $\bar{\mathbf{h}}_j$ are independent unit norm vectors. Therefore, we get
\begingroup\makeatletter\def\f@size{9}\check@mathfonts\def\maketag@@@#1{\hbox{\m@th\normalsize\normalfont#1}}
\begin{align}
\mathbb{E} \left[\textrm{SINR}_k^{\textrm{MF}}\right] \approx \frac{ \rho q }
{\frac{\rho \left(K-1 \right)}{K}+1}.\label{3222}
\end{align}\endgroup

\bibliographystyle{ieeetr}
\vspace{-8pt}
\bibliography{myrefs}

\begin{thebibliography}{10}

\bibitem{rusek2013scaling}
F.~Rusek, D.~Persson, B.~K. Lau, E.~G. Larsson, T.~L. Marzetta, O.~Edfors, and
  F.~Tufvesson, ``Scaling up {MIMO:} opportunities and challenges with very
  large arrays,'' {\em IEEE Sig. Process. Mag.}, vol.~30, no.~1, pp.~40--60,
  2013.

\bibitem{6736761}
E.~Larsson, O.~Edfors, F.~Tufvesson, and T.~Marzetta, ``Massive {MIMO} for next
  generation wireless systems,'' {\em IEEE Commun. Mag.}, vol.~52, no.~2,
  pp.~186--195, 2014.

\bibitem{marzetta2010noncooperative}
T.~L. Marzetta, ``Noncooperative cellular wireless with unlimited numbers of
  base station antennas,'' {\em IEEE Trans. Wireless Commun.}, vol.~9, no.~11,
  pp.~3590--3600, 2010.

\bibitem{brunobook}
B.~Clerckx and C.~Oestges, {\em {MIMO Wireless Networks: Channels, Techniques
  and Standards for Multi-Antenna, Multi-User and Multi-Cell Systems}}.
\newblock Academic Press, 2013.

\bibitem{5898372}
J.~Jose, A.~Ashikhmin, T.~Marzetta, and S.~Vishwanath, ``Pilot contamination
  and precoding in multi-cell {TDD} systems,'' {\em IEEE Trans. Wireless
  Commun.}, vol.~10, no.~8, pp.~2640--2651, 2011.

\bibitem{1327795}
B.~Hochwald, T.~Marzetta, and V.~Tarokh, ``Multiple-antenna channel hardening
  and its implications for rate feedback and scheduling,'' {\em IEEE Trans.
  Inf. Theory}, vol.~50, no.~9, pp.~1893--1909, 2004.

\bibitem{call}
C.~Neil, M.~Shafi, P.~Smith, and P.~Dmochowski, ``On the impact of antenna
  topologies for massive {MIMO} systems,'' {\em Proc. IEEE ICC}, 2015.

\bibitem{yang2013performance}
H.~Yang and T.~L. Marzetta, ``Performance of conjugate and zero-forcing
  beamforming in large-scale antenna systems,'' {\em IEEE J. Sel. Areas
  Commun.}, vol.~31, no.~2, pp.~172--179, 2013.

\bibitem{hoydis2013massive}
J.~Hoydis, S.~ten Brink, and M.~Debbah, ``Massive {MIMO} in the {UL/DL} of
  cellular networks: How many antennas do we need?,'' {\em IEEE J. Sel. Areas
  Commun.}, vol.~31, no.~2, pp.~160--171, 2013.

\bibitem{6377481}
S.-R. Lee, J.-S. Kim, S.-H. Moon, H.-B. Kong, and I.~Lee, ``Zero-forcing
  beamforming in multiuser {MISO} downlink systems under per-antenna power
  constraint and equal-rate metric,'' {\em IEEE Trans. Wireless Commun.},
  vol.~12, no.~1, pp.~228--236, 2013.

\bibitem{wagner2012large}
S.~Wagner, R.~Couillet, M.~Debbah, and D.~T. Slock, ``Large system analysis of
  linear precoding in correlated {MISO} broadcast channels under limited
  feedback,'' {\em IEEE Trans. Inf. Theory}, vol.~58, no.~7, pp.~4509--4537,
  2012.

\bibitem{6241389}
H.~Huh, G.~Caire, H.~Papadopoulos, and S.~Ramprashad, ``Achieving {massive
  MIMO} spectral efficiency with a not-so-large number of antennas,'' {\em IEEE
  Trans. Wireless Commun.}, vol.~11, pp.~3226--3239, 2012.

\bibitem{gao2011linear}
X.~Gao, O.~Edfors, F.~Rusek, and F.~Tufvesson, ``Linear pre-coding performance
  in measured {very-large} {MIMO} channels,'' in {\em Proc. IEEE Vehicular
  Technology Conference (VTC Fall)}, pp.~1 -- 5, 2011.

\bibitem{6608213}
K.~Truong and R.~Heath, ``Effects of channel aging in massive {MIMO} systems,''
  {\em J. Commun. and Net.}, vol.~15, no.~4, pp.~338--351, 2013.

\bibitem{1404883}
J.-C. Guey and L.~Larsson, ``Modeling and evaluation of {MIMO} systems
  exploiting channel reciprocity in {TDD} mode,'' in {\em Proc. IEEE Veh. Tech.
  Conf. (VTC)}, vol.~6, pp.~4265--4269, 2004.

\bibitem{choi2014}
J.~Choi, D.~Love, and P.~Bidigare, ``Downlink training techniques for {FDD}
  massive {MIMO} systems: Open-loop and closed-loop training with memory,''
  {\em IEEE J. Sel. Topics in Signal Process.}, vol.~8, no.~5, pp.~802--814,
  2014.

\bibitem{choi2013}
J.~Choi, Z.~Chance, D.~Love, and U.~Madhow, ``Noncoherent trellis coded
  quantization: A practical limited feedback technique for massive {MIMO}
  systems,'' {\em IEEE Trans. Commun.}, vol.~61, no.~12, pp.~5016--5029, 2013.

\bibitem{kuo2012compressive}
P.-H. Kuo, H.~Kung, and P.-A. Ting, ``Compressive sensing based channel
  feedback protocols for spatially-correlated massive antenna arrays,'' in {\em
  Proc. IEEE WCNC}, pp.~492--497, 2012.

\bibitem{au2011trellis}
C.~K. Au-Yeung, D.~J. Love, and S.~Sanayei, ``Trellis coded line packing{:}
  large dimensional beamforming vector quantization and feedback
  transmission,'' {\em IEEE Trans. Wireless Commun.}, vol.~10, no.~6,
  pp.~1844--1853, 2011.

\bibitem{love2003grassmannian}
D.~Love, R.~W. {Heath Jr.}, and T.~Strohmer, ``Grassmannian beamforming for
  multiple-input multiple-output wireless systems,'' {\em IEEE Trans. on Inf.
  Theory}, vol.~49, no.~10, pp.~2735--2747, 2003.

\bibitem{love2008overview}
D.~J. Love, R.~W. {Heath Jr.}, V.~K.~N. Lau, D.~Gesbert, B.~D. Rao, and
  M.~Andrews, ``An overview of limited feedback in wireless communication
  systems,'' {\em IEEE J. Sel. Areas Commun.}, vol.~26, no.~8, pp.~1341--1365,
  2008.

\bibitem{au2007performance}
C.~K. Au-Yeung and D.~J. Love, ``On the performance of random vector
  quantization limited feedback beamforming in a {MISO} system,'' {\em IEEE
  Trans. Wireless Commun.}, vol.~6, no.~2, pp.~458--462, 2007.

\bibitem{raghavan2007systematic}
V.~Raghavan, R.~Heath, and A.~M. Sayeed, ``Systematic codebook designs for
  quantized beamforming in correlated {MIMO} channels,'' {\em IEEE J. Sel.
  Areas Commun.}, vol.~25, no.~7, pp.~1298--1310, 2007.

\bibitem{kim2011mimo}
T.~Kim, D.~Love, and B.~Clerckx, ``{MIMO} systems with limited rate
  differential feedback in slowly varying channels,'' {\em IEEE Trans.
  Commun.}, vol.~59, no.~4, pp.~1175--1189, 2011.

\bibitem{choi2012new}
J.~Choi, B.~Clerckx, N.~Lee, and G.~Kim, ``A new design of polar-cap
  differential codebook for temporally/spatially correlated {MISO} channels,''
  {\em IEEE Trans. Wireless Commun.}, vol.~11, no.~2, pp.~703--711, 2012.

\bibitem{love2006limited}
D.~J. Love and R.~W. Heath~Jr, ``Limited feedback diversity techniques for
  correlated channels,'' {\em IEEE Trans. Veh. Tech.}, vol.~55, no.~2,
  pp.~718--722, 2006.

\bibitem{dMirzaLimited}
J.~Mirza, P.~Dmochowski, P.~Smith, and M.~Shafi, ``Limited feedback multiuser
  {MISO} systems with differential codebooks in correlated channels,'' in {\em
  Proc. IEEE Int. Conf. on Commun.}, pp.~3979--3984, 2013.

\bibitem{6648514}
J.~Mirza, P.~Dmochowski, P.~Smith, and M.~Shafi, ``A differential codebook with
  adaptive scaling for limited feedback {MU MISO} systems,'' {\em IEEE Wireless
  Commun. Lett.}, vol.~3, no.~1, pp.~2--5, 2014.

\bibitem{6824239}
J.~Mirza, P.~Smith, M.~Shafi, P.~Dmochowski, A.~Firag, and A.~Papathanassiou,
  ``Double-cap differential codebook structure for {MU MISO} systems in
  correlated channels,'' {\em IEEE Wireless Commun. Lett.}, pp.~441--444, 2014.

\bibitem{marcellin1990trellis}
M.~W. Marcellin and T.~R. Fischer, ``Trellis coded quantization of memoryless
  and {Gauss-Markov} sources,'' {\em IEEE Trans. Commun.}, vol.~38, no.~1,
  pp.~82--93, 1990.

\bibitem{ungerboeck1982channel}
G.~Ungerboeck, ``Channel coding with multilevel/phase signals,'' {\em IEEE
  Trans. Inf. Theory}, vol.~28, pp.~55--67, 1982.

\bibitem{forney1973viterbi}
G.~D. Forney~Jr, ``The {V}iterbi algorithm,'' {\em IEEE Proc.}, vol.~61, no.~3,
  pp.~268--278, 1973.

\bibitem{jindal2006mimo}
N.~Jindal, ``{MIMO} broadcast channels with finite-rate feedback,'' {\em IEEE
  Trans. Inf. Theory}, vol.~52, no.~11, pp.~5045--5060, 2006.

\bibitem{yoo2006optimality}
T.~Yoo and A.~Goldsmith, ``On the optimality of multiantenna broadcast
  scheduling using zero-forcing beamforming,'' {\em IEEE J. Sel. Areas
  Commun.}, vol.~24, no.~3, pp.~528--541, 2006.

\bibitem{1468466}
M.~Joham, W.~Utschick, and J.~Nossek, ``Linear transmit processing in {MIMO}
  communications systems,'' {\em IEEE Trans. Sig. Process.}, vol.~53, no.~8,
  pp.~2700--2712, 2005.

\bibitem{godara1997application}
L.~C. Godara, ``Application of antenna arrays to mobile communications, {P}art
  {II}. {B}eam-forming and direction-of-arrival considerations,'' {\em Proc.
  IEEE}, vol.~85, no.~8, pp.~1195--1245, 1997.

\bibitem{peel2005vector}
C.~B. Peel, B.~M. Hochwald, and A.~L. Swindlehurst, ``A vector-perturbation
  technique for near-capacity multiantenna multiuser communication-{P}art {I}:
  channel inversion and regularization,'' {\em IEEE Trans. Commun.}, vol.~53,
  no.~1, pp.~195--202, 2005.

\bibitem{Hochwald02space-timemultiple}
B.~Hochwald and S.~Vishwanath, ``Space-time multiple access: Linear growth in
  the sum rate,'' in {\em Proc. 40th Annual Allerton Conf. Communications,
  Control and Computing}, 2002.

\bibitem{5510182}
L.~Yu, W.~Liu, and R.~Langley, ``{SINR} analysis of the subtraction-based {SMI}
  beamformer,'' {\em IEEE Trans. Signal Process.}, vol.~58, no.~11,
  pp.~5926--5932, 2010.

\bibitem{newman1979principles}
W.~M. Newman and R.~F. Sproull, {\em Principles of {I}nteractive {C}omputer
  {G}raphics}.
\newblock McGraw-Hill, 1979.

\bibitem{kotz1994continuous}
N.~L. Johnson, S.~Kotz, and N.~Balakrishnan, {\em Continuous {U}nivariate
  {D}istributions}, vol.~1.
\newblock John Wiley \& Sons, 1994.

\bibitem{951380}
S.~Loyka, ``Channel capacity of {MIMO} architecture using the exponential
  correlation matrix,'' {\em IEEE Commun. Lett.}, vol.~5, pp.~369--371, 2001.

\bibitem{1285039}
J.-S. Jiang and M.~Ingram, ``Distributed source model for short-range {MIMO},''
  in {\em Proc. IEEE Veh. Tech. Conf.}, vol.~1, pp.~357--362, 2003.

\bibitem{abouda2006effect}
A.~A. Abouda, H.~M. El-Sallabi, and S.~H{\"a}ggman, ``Effect of antenna array
  geometry and {ULA} azimuthal orientation on {MIMO} channel properties in
  urban city street grid,'' {\em Progress In Electromagnetics Research},
  vol.~64, pp.~257--278, 2006.

\bibitem{tufvesson_mimo}
X.~Gao, O.~Edfors, F.~Rusek, and F.~Tufvesson, ``Massive {MIMO} performance
  evaluation based on measured propagation data,'' {\em submitted for
  publication}, [Online]. Available: http://arxiv.org/abs/1403.3376,.

\bibitem{ist2007deliverable}
{IST-WINNER II Deliverable 1.1.2 v.1.2}, ``{WINNER} {II} channel models,'' {\em
  {IST-WINNER2, Technical Report}}, 2007.

\end{thebibliography}

\end{document}